**Title: Modelling and docking of Indian SARS-CoV-2 spike protein 1 with ACE2: implications for co-morbidity and therapeutic intervention**

Dhanashree Jagtap[a*], Selvaa Kumar C[b*], Smita Mahale[a] , Vainav Patel[a#]

a. ICMR- National Institute for Research in Reproductive Health, Jehangir Merwanji Street, Parel, Mumbai 400012, Maharashtra, India
b. School of Biotechnology and Bioinformatics, D Y Patil Deemed to be University, CBD Belapur, Navi Mumbai, Maharashtra, India

*Both the authors have contributed equally to this manuscript.

#Corresponding Author: Vainav Patel, PhD; Email: patelv@nirrh.res.in

**Abstract**

Presently, India bears amongst the highest burden of non-communicable diseases such as diabetes mellitus (DM), hypertension (HT), and cardio vascular disease (CVD) and thus represents a vulnerable target to the SARS-CoV-2/COVID-19 pandemic. Involvement of the angiotensin converting enzyme 2 (ACE2) in susceptibility to infection and pathogenesis by SARS-CoV-2 is currently an actively pursued research area. An increased susceptibility to infection in individuals with DM, HT and CVD together with higher levels of circulating ACE2 in these settings presents a scenario where interaction with soluble ACE2 may result in disseminated virus-receptor complexes that could enhance virus acquisition and pathogenesis. Thus, understanding the SARS-CoV-2 receptor binding domain-ACE2 interaction, both membrane bound and in the cell free context may contribute to elucidating the role of co-morbidities in increased susceptibility to infection and pathogenesis. Both Azithromycin and Hydroxychloroquine (HCQ) have shown efficacy in mitigating viral carriage in infected individuals. Furthermore, each of these compounds generate active metabolites which in turn may also modulate virus-receptor interaction and thus influence clinical outcomes. In this study, we model the structural interaction of S1 with both full-length and soluble ACE2. Additionally, therapeutic drugs and their active metabolites were docked with soluble ACE2 protein. Our results show that S1 from either of the reported Indian sequences can bind both full-length and soluble ACE2, albeit with varying affinity that can be attributed to a reported substitution in the RBD. Furthermore, both Azythromycin and HCQ together with their active metabolites can allosterically affect, to a range of extents, binding of S1 to ACE2.

## Introduction

Coronavirus disease 2019 (COVID-19) caused by the novel severe acute respiratory syndrome coronavirus 2 (SARS-CoV-2) pandemic has impacted around 1.7 million people with more than 10000 deaths from 210 countries across the globe so far (Zhu N et al, 2020; ; CSSE, Johns Hopkins University). As of 11th April 2020, India has 6565 active cases, 239 deaths with 642 individuals who have been documented to have recovered (Ministry of Health and Family Welfare, Government of India; https://www.mohfw.gov.in/). All these figures are expected to escalate rapidly. Several clinical and epidemiological studies reporting on COVID-19 have linked comorbidities such as diabetes (DM), hypertension (HT) and cardiovascular disease (CVD) with increased risk of infection leading to increased mortality (Guan W et al, 2020; Fang L et al, 2020). Angiotensin converting enzyme 2 (ACE2), the receptor for SARS-CoV-2, is widely expressed by several organs including the oral mucosa, lung, intestine, kidney, blood vessels, and on immunoreactive cells (Zou X et al, 2020). Elevated levels of ACE2 could potentially result in increased susceptibility to infection by SARS-CoV-2 (Xu H et al, 2020). Patients with DM, HT and CVD, which seem to be the most common comorbidities in patients with COVID-19, are typically treated with drugs that inhibit the renin-angiotensin system (RAS), including angiotensin-converting enzyme inhibitors (ACEIs) or angiotensin receptor blockers (ARBs). RAS inhibitors are very commonly used not only for management of blood pressure but also for protection from disease-associated inflammation and organ remodelling (Simoes ESAC et al, 2016). A consequence of this therapy, increased levels of ACE2, including its cell-free soluble form are observed in these individuals (Vuille-dit-Bille RN et al, 2015; Hoffmann M et al, 2020). The high proportion of patients with severe COVID-19 and these comorbidities prompted us to probe the molecular mechanism by which SARS-CoV-2 attaches and interacts with membrane bound and cell free forms of ACE2.

At present, a combination of two known drugs namely Azithromycin and Hydroxychloroquine (HCQ) are promising therapeutic interventions used for management of COVID-19 patients (Gautret P et al, 2020). Direct and potentially competitive effects of these drugs with virus-receptor interaction have heretofore been unexplored. Thus, in this study we investigated the potential interaction of the drugs Azithromycin and HCQ together with their active metabolites and biologically relevant forms of ACE2 to provide structural insights that would help in understanding the increased pathogenesis in these high risk individuals.

## Materials and Methods

### Multiple sequence alignment

SARS-CoV-2spike protein 1 (S1) for Indian sequence 1 (INS1; Yadav P, et al., 2020; Acc. No. MT050493), 2 (INS2; Yadav P, et al., 2020; Acc. No. MT012098.1), Wuhan SARS-CoV-2 (Wu F, et al., 2020; Acc. No. QHD43416) and SARS-CoV ( He R, et al., 2018; NP-828851.1) were downloaded from NCBI database for multiple sequence alignment to identify potential indels and substitutions between these four protein sequences using CLUSTAL-Omega tool (Fabian Sievers et al, 2011). We performed two multiple sequence alignments; the first alignment comprised of INS1, INS2 and Wuhan SARS-CoV-2 and the second multiple sequence alignment comprised of INS1, INS2, Wuhan SARS-CoV-2 and SARS-CoV sequences.

### Homology modelling

The ACE2 protein of *Homo sapiens* was downloaded from Uniprot database with an Accession Number: Q9BYF1 (Bairoch A, 1996). This protein sequence is made up of 805 amino acids (aa) which is classified into extracellular domain (aa 18-740); helical domain (aa 741-761) and cytoplasmic domain (aa 792-805). In particular, the interface regions aa 30-41; aa 82-84 and aa 353-357 were preferred by S1 for protein-protein interaction. The circulating soluble form of ACE2 (sACE2) without the membrane anchor has been reported in blood (Batlle D et al, 2020). Apart from the full length ACE2, sACE2 without the transmembrane region with residues ranging from 18-708 has also been reported in PTM/Processing- Uniprot database. Protein Data Bank (PDB) search was done for identification of potential structural templates for both complete and sACE2. The crystal structure of ACE2 was available in PDB (PDB ID: 6m18) but with many missing residues (Berman H, 2000; Yan R et al, 2020). Therefore, homology modelling of the ACE2 protein sequence of *Homo sapiens* using SWISS-MODEL server was opted (Waterhouse A et al, 2018). From the modelled ACE2 protein, the complete and sACE2 was generated for further analysis. The two reported S1 sequences for Indian SARS-CoV-2 (INS1 and INS2) were downloaded from NCBI. Both these sequences were modelled using SWISS-MODEL online server. Finally, modelled ACE2 and S1 from INS1 and INS2 were considered for structure validation using PROCHECK-Ramachandran plot server (Laskowski R A et al, 1993). Modelled ACE2, INS1 and INS2 3D structures were energy minimized using CHIMERA software (Pettersen E et al, 2004).

## Protein-Protein docking

Both complete and sACE2 were considered for Protein-Protein docking with receptor binding domain (RBD) from S1 of INS1 and INS2 using HADDOCK server (G.C.P van Zundert et al, 2016). For docking purpose, the active and passive residues were obtained through literature survey. Active residues associated with ACE2 included residues within the range of aa 30-41, aa 82-84 and aa 353-357 which were obtained from Uniprot Database.

Regarding the active residues for S1, the complete RBD was considered which included residues from aa 319-541. Within this region lies the receptor binding motif (RBM) (aa 437-508). The residues critically involved in receptor interaction were identified through literature review (Tai W et al, 2020; Lan J et al, 2020). Passive residues were defined automatically through the check box for ACE2 and S1 from INS1 and INS2. In total, four docking experiments were performed which comprised of complete ACE2 with S1 from INS1 and INS2 and sACE2 with S1 from INS1 and INS2. For every run, HADDOCK software generated 10 clusters of four poses each, along with the HADDOCK score, buried surface area and energy details. Of these, the cluster with least HADDOCK score was considered for their binding energy analysis using PRODIGY server for all the four docking experiments (Vangone A et al, 2019). Furthermore, the docked poses were generated using CHIMERA software (Pettersen E et al, 2004). Hydrogen bonding analysis was carried out using Discovery Studio Visualizer (Discovery Studio 2019).

## Protein Ligand docking

Modelled ACE2 was docked with drugs retrieved from PubChem (Kim S et al, 2019), MOLBASE (www.molbase.com) and sketched structures. Chemical structures for Azithromycin (PubChem ID: 55185), Descladinose-azithromycin (PubChem ID: 71315587), Desethyl chloroquine (DCQ) (PubChem ID:95478), Desethyl hydroxychloroquine (PubChem ID:71826) and Hydroxychloroquine (HCQ) (PubChem ID:3652) were downloaded from PubChem database. 9a-N-desmethyl azithromycin with CAS NO: 76801-85-9 was downloaded from MOLBASE database, and bis desethyl hydroxychloroquine (BDCQ) was sketched using PUBCHEM Sketcher 2.4 (Ihlenfeldt et al., 2009). All the seven drug metabolites were considered for stable conformer generation using FRee Online druG conformation generation (FROG2) online server (Maria A et al, 2020). The input drug description used was SMILES (simplified molecular-input line-entry system) string and the output format was PDB with a single product while the rest of the calculation parameters were set at default. AutoDock Tools 1.5.6 was used for protein-ligand docking (Morris, G et al, 2009).

Based on the literature review, the active site residues for ACE2 were Arg273, His345, Pro346, Glu375, His505 and Tyr515 (Tai W et al, 2020). Before docking, Kollman charges and Gasteiger charges were added to the protein and ligands respectively. The grid was centered within these six residues that form the active site. The size of the grid box was 70 Å, 64 Å and 80Å for x, y and z respectively. Furthermore, the grid center was set to 2.250, 0.944 and 7.750 for x, y and z respectively. AutoGrid 4.0 and AutoDock 4.0 programs were used to generate grid maps. The best ten conformers were generated using Lamarckian Genetic Algorithm. The binding energy and inhibition constant for each pose was calculated and the best selected poses were visualized using Discovery Studio Visualizer (Discovery Studio 2019).

## Results

### Multiple Sequence alignment

S1 from INS1, INS2 and Wuhan SARS-CoV-2 protein sequences were considered for multiple sequence alignment using Clustal-Omega. The overall identity between S1 of INS1 and INS2 was 99.84%. The overall identity between INS1 and Wuhan SARS-CoV-2 was 99.92 %. Tyr145 from INS1 and Wuhan SARS-CoV-2 is deleted in INS2 within the N-terminal domain (NTD). Similarly, Arg408 from INS1 and Wuhan SARS-CoV-2 is substituted by Ile in INS2 (at position 407). Here, a positively charged residue is substituted by a hydrophobic residue within the RBD. Also, at position 929 in INS2 Val is substituted by Ala in INS1 and Wuhan SARS-CoV-2. In total, two substitutions and a single deletion was observed between these two sequences (Figure 1). Also, INS1, INS2 and Wuhan SARS-CoV-2 sequences were compared with SARS-CoV. Here we observed changes in the N-terminal domain and Receptor Binding Domain (Supplementary Figure 1).

### Homology modelling

ACE2 sequence was considered as a query sequence in SWISS-MODEL to identify potential template. The crystal structure with PDB ID: 6m18 chain B was listed as a potential template with a query coverage of 99 % and 100 % amino acid identity. The generated model of ACE2 had aa 21-768 which had both extracellular and helical domains. This complete model was validated using Ramachandran plot wherein the most favoured region was 91.4%; additional allowed region was 8.2%; generously allowed region was 0.4% and disallowed region was 0% which confirmed that the structure was very stable (Figure 2a). The whole model had aa 21-768 which is the combination of extracellular and helical domain of which, only the extracellular domain was retained (aa 21-740) by removing the helical domain which forms the complete ACE2 model. Furthermore, processed ACE2 was isolated (aa 21-708) to generate the sACE2 (Figure 3a).

For S1, 6VSB chain A was listed as potential template with an amino acid identity of 99.17 % and a query coverage of 95% (Wrapp, D et al, 2020). All the domain details were obtained from literature review (Xia S et al, 2020). This generated model comprised of aa 27-1146 which forms complete S1 (aa 13-541) and partial S2 region (aa 778-1213) wherein the N-terminal domain (NTD) was from aa 13-305; RBD from aa 319-436 and aa 509-541; RBM from aa 437-508; fusion peptide from aa 788-806 and the heptad repeat region1 from aa 912-984; the heptad repeat region 2 from aa 1163-1213 could not be modelled (Figure 3b). As per Ramachandran Plot, the most favoured region was 86.1%; additional allowed region was 11.9%; generously allowed region was 1.7 and disallowed region was 0.3% (Figure 2b). When the RBD binding site was evaluated in INS1 and INS2 we observed that Arg 408 of INS1 interacts with Asp405 and Thr376 resulting in a salt bridge and hydrogen bond within the RBD respectively. Interestingly, both these interactions are lost post substitution of Arg with Ile in INS2 (Figure 3c).

**Protein-Protein docking**

ACE2 was docked with the RBD of INS1 and INS2 which includes the RBM (described in Methods). The top cluster generated by HADDOCK was considered to reliable based on available literature. Higher HADDOCK dock score was observed between ACE2 and INS1 compared to ACE2 and INS2. The buried surface area of INS1 was higher than INS2. Three salt bridges along with eleven hydrogen bonds were observed between complete ACE2 and INS1 (Figure 4a, b and Table 1). Furthermore, we observed the overall interface region of ACE2 which comes in close contact with INS1 and INS2. Interestingly, within ACE2 charged residues seem to play a critical role in these interactions. Of the ten interactions observed, seven were negatively charged while three were positively charged (Table 2). Next, we analysed the interaction between complete ACE2 and INS2. There were three salt bridges observed between INS2 and ACE2 along with twelve hydrogen bonds. Within the interface of ACE2 there were five negatively charged and five positively charged residues which come in direct contact with INS2 (Figure 4b and Table 2). Docking of sACE2 with INS1 and INS2 (Figure 4c and d) showed that HADDOCK score of sACE2 with INS1 was higher than that of INS2 (Table 1). An increase in buried surface area was observed in INS1 compared to INS2. To ascribe biological relevance to the HADDOCK score it was further processed with PRODIGY online server to calculate binding energy. Reassuringly, we observed that HADDOCK scores were in concordance with the PRODIGY scores (Table 1).

Further, interaction analysis between sACE2 and INS1 (Table 2) revealed two salt bridges and fourteen hydrogen bonds between INS1 and sACE2. The interface had seven negatively charged and three positively charged residues in direct contact. For interaction between sACE2 with INS2, two salt bridges and eleven hydrogen bonds were observed. Within sACE2, there were four negatively charged and three positively charged residues that were involved in interaction with INS2 (Table 2) (Figure 4c, d). Surface based analysis of ACE2 revealed that it has a relatively larger hydrophilic region and a smaller hydrophobic region available for S1 interaction (Figure 5).

**Protein -ligand docking**

In order to evaluate the binding affinity of actively considered therapeutic drugs and their active metabolites in the context of ACE2-S1 binding pocket, a local docking approach was employed. Azithromycin, Descladinose-azithromycin, 9a-N-desmethyl-azithromycin, Hydroxychloroquine (HCQ), Desethyl chloroquine, Desethyl hydroxychloroquine, and bis desethyl hydroxychloroquine were docked with ACE2 protein within the active site. Azithromycin and its active metabolites Descladinose-azithromycin and 9a-N-Desmethyl-azithromycin showed better drug binding and inhibition constants compared to HCQ and its active metabolites (Figure 6a-c and Table 3). Amongst HCQ and its active metabolites, only bis desethyl hydroxychloroquine showed comparable (to Azithromycin) drug binding and inhibition constant (Figure 6d-g and Table 3). Negatively charged amino acids (on ACE2) Asp350, Asp382, Glu402 and Glu375 were preferred for all drug molecules except for HCQ which showed preference for positively charged amino acids (Arg393 and His401). Considering the proximity of the S1 binding site to the drug binding pocket of these compounds (Figure 3a), our results indicate the possibility of an allosteric modulation by these drugs.

## Discussion

In this study, we provide structural insights into the interaction between S1 and both soluble as well as full-length human ACE2 using the only two available viral sequences from India. Also, we elucidate the influence of potential therapeutics - Azythromycin and HCQ, together with their active metabolites, on the binding of S1 with ACE2. To begin with, we confirmed with Indian SARS-CoV-2 sequences that a stronger interaction with ACE2 exists compared to the SARS-CoV with ensuing implications for higher infection rates and transmission as has been reported recently (Chen Y, et al., 2020). A preliminary analysis of the two S1 sequences from India, that has witnessed introductions of the virus into the population from geographically distinct sources, reveals two substitutions and a single deletion. Interestingly, one of these substitutions is at position aa 407 within the RBD where a positively charged residue (Arg) is substituted by a hydrophobic residue (Ile). Other than obvious implications for virus-ACE2 binding, we investigated whether this substitution would also affect soluble ACE2 and S1 interaction. Our results show that INS1, most similar to the prototypical Chinese sequence (Supplementary Figure 1), had greater affinity for both forms of ACE2 compared to INS2. Intriguingly, INS2 and not INS1, seemed to show preferential affinity for sACE2. Further, the difference in affinities was lower when bound to sACE2. This has implications from both the co-morbidity as well as interventional perspective. Our results suggest that S1-ACE2 complexes are stable and could potentially be systemically disseminated, more so in individuals with DM, HT and CV, resulting in higher co-morbidity and susceptibility to infection. Any vaccine generating neutralizing antibodies, for example, would need to disrupt these soluble complexes as well and conversely, if successful in doing so, would dramatically reduce susceptibility in vulnerable populations.

The second major thrust of our study was to understand how potential therapeutics, being actively considered for widespread prophylactic and therapeutic intervention would affect virus-receptor interaction. Considering that an earlier report for SARS-CoV had suggested the possibility of such modulation (Martin JV, et al., 2005), we expanded our analysis to include active metabolites of these drugs. The drugs/metabolites evaluated were Azithromycin, Descladinose-azithromycin, 9a-N-desmethyl-azithromycin, Hydroxychloroquine, Desethyl chloroquine, Desethyl hydroxychloroquine, and bis desethyl hydroxychloroquine (BDCQ). We report for the first time that each of these compounds have the potential to allosterically modulate S1-ACE2 binding when studied using a protein-ligand approach. HCQ and its metabolites with the exception of BDCQ showed lower affinity and higher inhibitory constants compared to those of Azithromycin. Significantly, metabolites in general showed similar binding affinity compared to their respective parent compounds. In fact, in the case of Hydroxychloroquine, BDCQ showed almost 30-fold better binding to ACE2 and this affinity was similar to that observed for Azithromycin. The lack of sequence information both globally and especially from India (Forster P, et al., 2020; Yadav P, et al., 2020) has limited our ability to robustly examine true consensus sequences for our analysis. Also, our insights need to be experimentally validated *in vitro* and *in vivo*. In conclusion, our results highlight a putative role of S1-sACE2 complexes in viral pathogenesis in susceptible populations. Further, we posit a direct role for ACE2 interacting drugs and other lead compounds (Goswami D, et al., 2020) in preventing infection, ameliorating viral carriage and eventually affecting transmission.

**Acknowledgement**

The authors would like to acknowledge intramural support from ICMR-NIRRH and the School of Biotechnology and Bioinformatics, D.Y. Patil Deemed to be University for providing access to the softwares for biological data analysis. This work was not funded by any external funding agencies.

**Competing interests**

None

**Author Contributions:**

DJ and SKC: Contributed equally to this manuscript. They analysed data, generated figures and helped in writing with the manuscript. SM: Performed data analysis and reviewed the manuscript. VP, DJ and SKC: Conceived manuscript performed data analysis and wrote the manuscript.

## Figure legends

**Figure 1.** Multiple sequence alignment of SARS-CoV-2 spike glycoprotein (S1) of Indian sequence 2 (MT012098.1), Indian sequence 1 (MT050493) and SARS-CoV2-WUHAN-QHD43416. The Wuhan SARS-CoV-2 is sequentially identical to INS1 at N-terminal domain (NTD) and the RBD. Wuhan SARS-CoV-2 and INS2 is only identical at position 929 wherein valine is substituted by alanine. Comparing INS1 with INS2, a single indel was observed at position 145. Substitution of charged residue with hydrophobic residue was observed at position 407. N-terminal Domain, RBD and RBM domains are highlighted in boxes. Deletion of tyrosine and substitution of arginine with isoleucine is highlighted by a downward arrow.

**Figure 2.** The Ramachandran plot for the modelled proteins (a) ACE2 protein of Homo sapiens and (b) for SARS-CoV-2 Indian spike protein (S1).

**Figure 3.** Homology model of human ACE2 and Indian SARS-CoV2 protein (a) Modelled ACE2 protein with domain details shown in ribbon file format. The extracellular domain comprises of green and pink colour. The helical domain is highlighted (blue) along with the soluble ACE2 protein (Dark green). SARS-COV-2 binding site is shown in surface format (cyan) and the drug binding site in sphere format (magenta) (b). The modelled Indian SARS-COV-2 Spike glycoprotein S1 with the N-terminal domain (Yellow), RBD (Blue) and RBM (Green), fusion peptide (Cyan) and Heptad repeat 1 (Magenta) is displayed using ribbon format. (c) RBD binding site INS1 showing salt bridges of Arg 408 with Asp405 and Thr376 Both these interactions are lost post substitution of Arg with Ile in INS2.

**Figure 4.** Protein-protein docking (a) SARS-CoV-2 Indian sequence 1 with complete ACE2, (b) SARS-CoV-2 Indian sequence 2 with complete ACE2, (c) Indian sequence 1 with soluble ACE2 and (d) Indian sequence 2 with soluble ACE2. The orange colour represents the Indian SARS-CoV-2 protein and dark green colour represents ACE2 protein.

**Figure 5.** Surface view of ACE2 protein depicting the protein-protein binding site. The blue colour depicts the hydrophilic region and the red depicts the hydrophobic region.

**Figure 6.** Protein-ligand docking of soluble ACE2 with (a) Azithromycin, (b) Descladinose-azithromycin, (c) 9a-N-Desmethyl-azithromycin, (d) Hydroxychloroquine, (e) Desethyl chloroquine, (f) Desethyl hydroxychloroquine, (g) Bisdesethyl hydroxychloroquine. The soluble ACE2 protein is shown in ribbon format in green colour, sidechains are shown in magenta and the drug in orange colour.

**Supplementary Figure 1.** Multiple sequence alignment of SARS-CoV-2 spike glycoprotein (S1) of SARS-CoV-NP_828851.1 (SARS-CoV-2003), Indian sequence 2 (MT012098.1), Indian sequence 1 (MT050493) and SARS-CoV2-WUHAN-QHD43416.

**Figures**

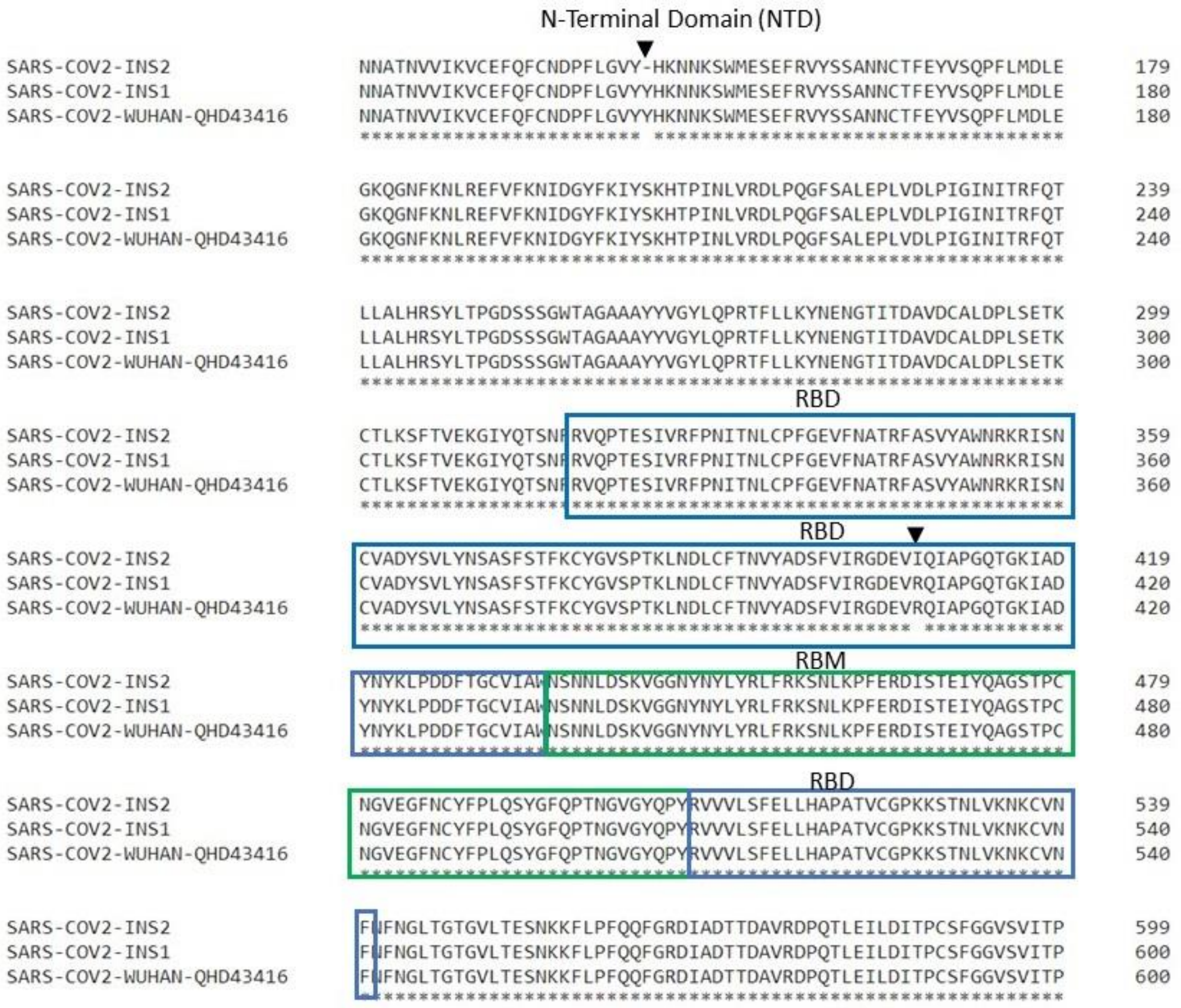

**Figure 1.** Multiple sequence alignment of SARS-CoV-2 spike glycoprotein (S1) of Indian sequence 2 (MT012098.1), Indian sequence 1 (MT050493) and SARS-CoV2-WUHAN-QHD43416. The Wuhan SARS-CoV-2 is sequentially identical to INS1 at N-terminal domain (NTD) and the RBD. Wuhan SARS-CoV-2 and INS2 is only identical at position 929 wherein valine is substituted by alanine. Comparing INS1 with INS2, a single indel was observed at position 145. Substitution of charged residue with hydrophobic residue was observed at position 407. N-terminal Domain, RBD and RBM domains are highlighted in boxes. Deletion of tyrosine and substitution of arginine with isoleucine is highlighted by a downward arrow.

2(a)

2(b)

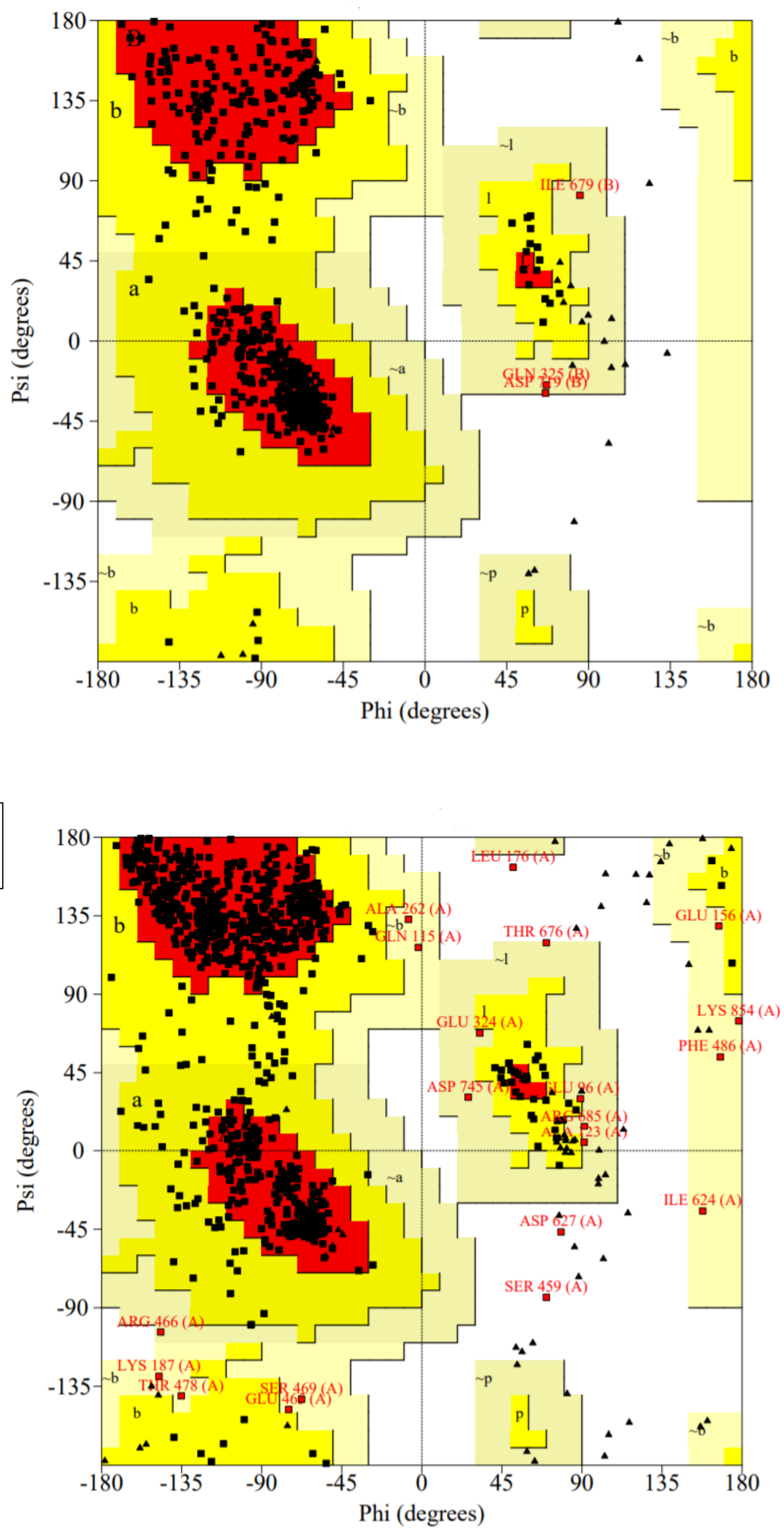

**Figure 2.** The Ramachandran plot for the modelled proteins (a) ACE2 protein of Homo sapiens and (b) for SARS-CoV-2 Indian spike protein (S1).

3(a)

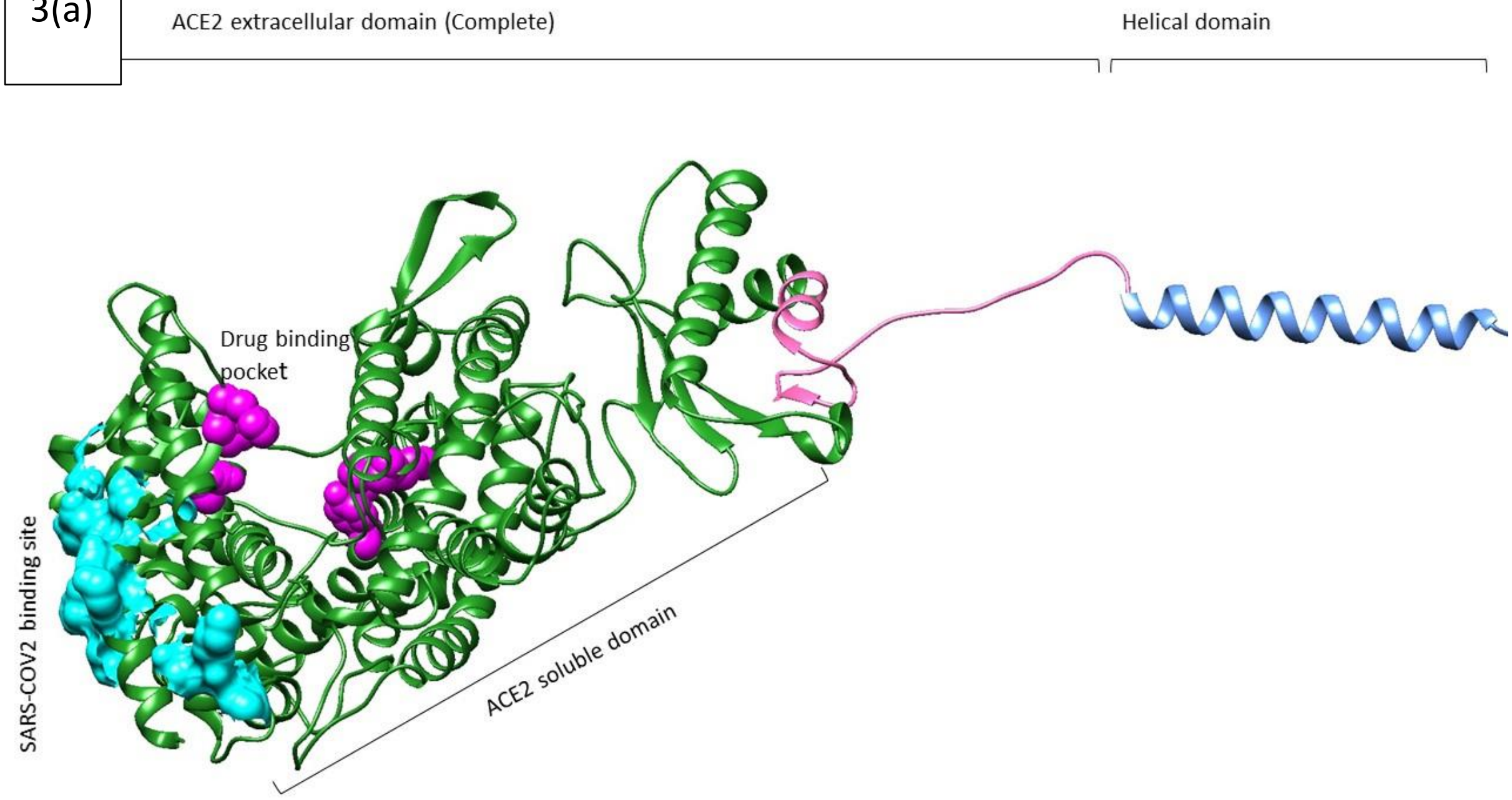

ACE2 extracellular domain (Complete)
Helical domain
Drug binding pocket
SARS-COV2 binding site
ACE2 soluble domain

3(b)

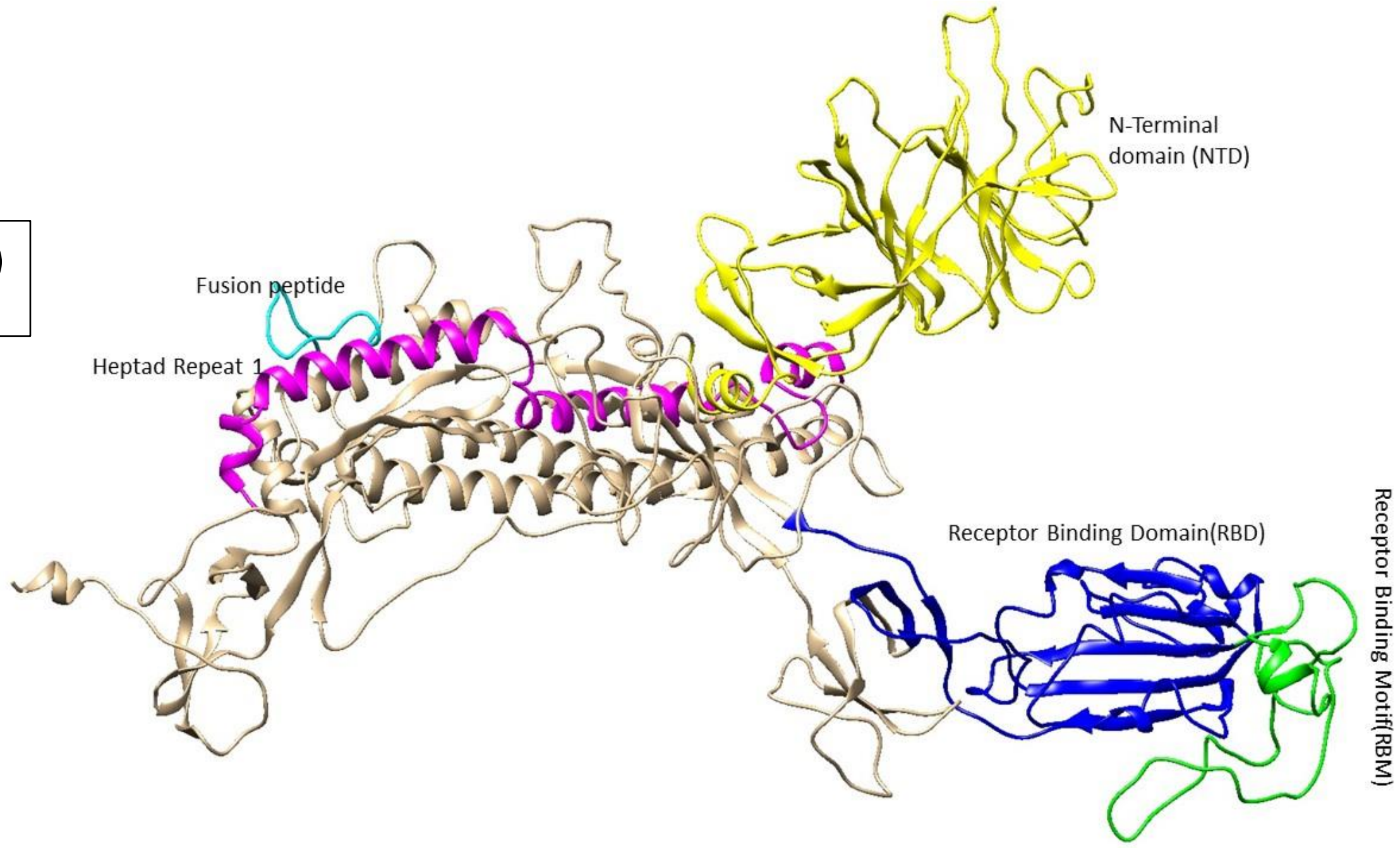

N-Terminal domain (NTD)
Fusion peptide
Heptad Repeat 1
Receptor Binding Domain(RBD)
Receptor Binding Motif(RBM)

3(c)

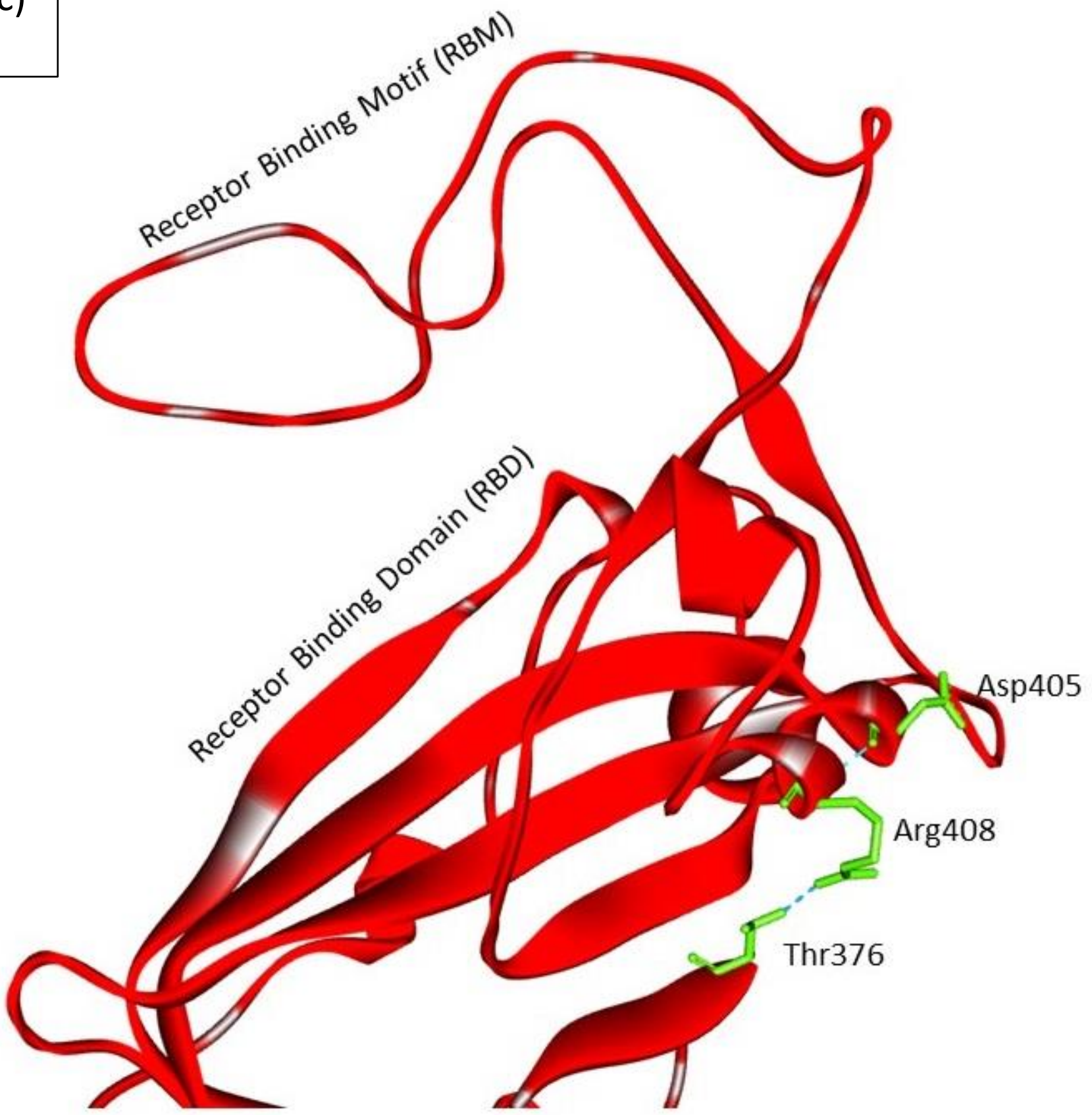

**Figure 3.** Homology model of human ACE2 and Indian SARS-CoV2 protein (a) Modelled ACE2 protein with domain details shown in ribbon file format. The extracellular domain comprises of green and pink colour. The helical domain is highlighted (blue) along with the soluble ACE2 protein (Dark green). SARS-COV-2 binding site is shown in surface format (cyan) and the drug binding site in sphere format (magenta) (b). The modelled Indian SARS-COV-2 Spike glycoprotein S1 with the N-terminal domain (Yellow), RBD (Blue) and RBM (Green), fusion peptide (Cyan) and Heptad repeat 1 (Magenta) is displayed using ribbon format. (c) RBD binding site INS1 showing salt bridges of Arg 408 with Asp405 and Thr376 Both these interactions are lost post substitution of Arg with Ile in INS2.

4(a)

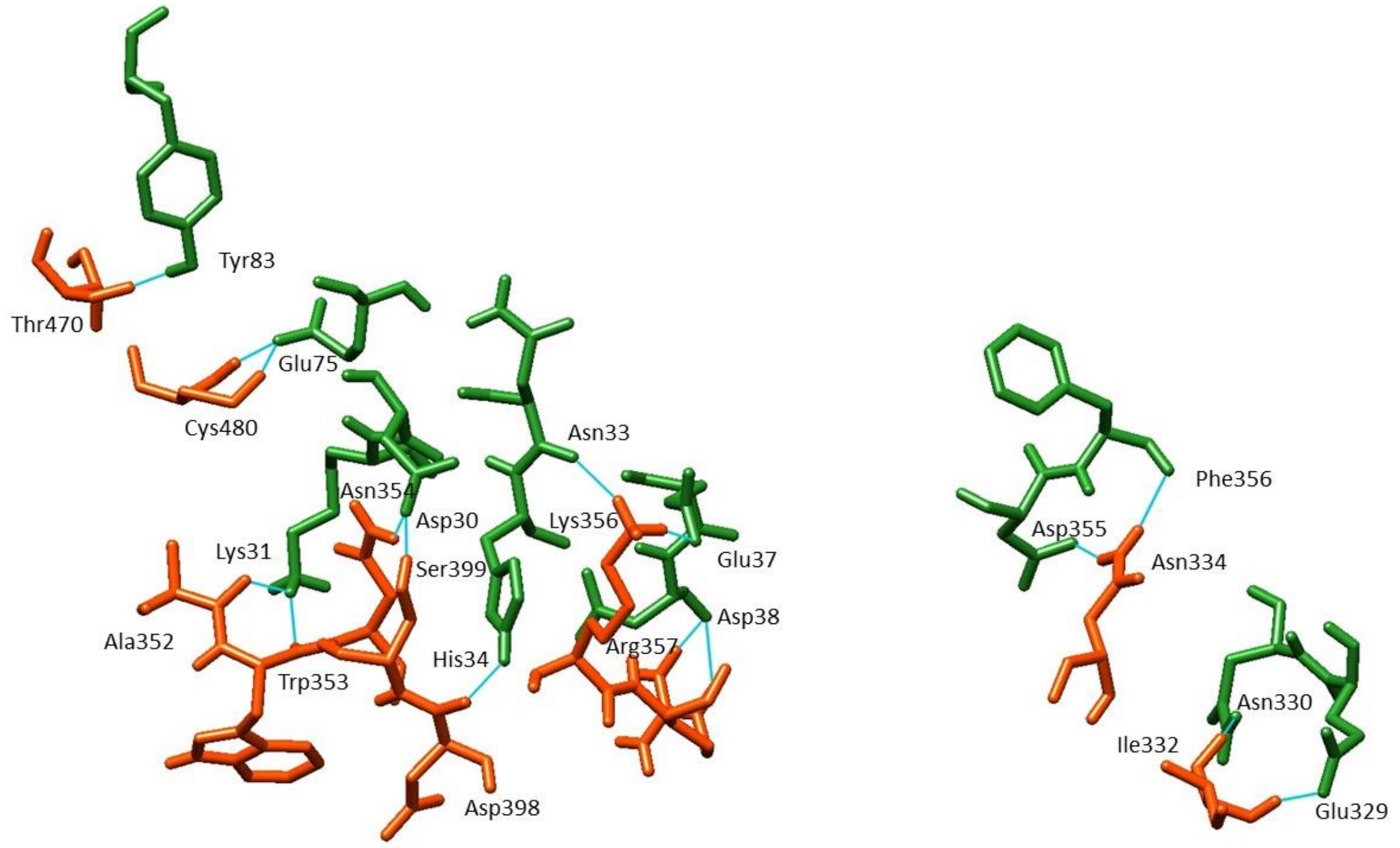

4(b)

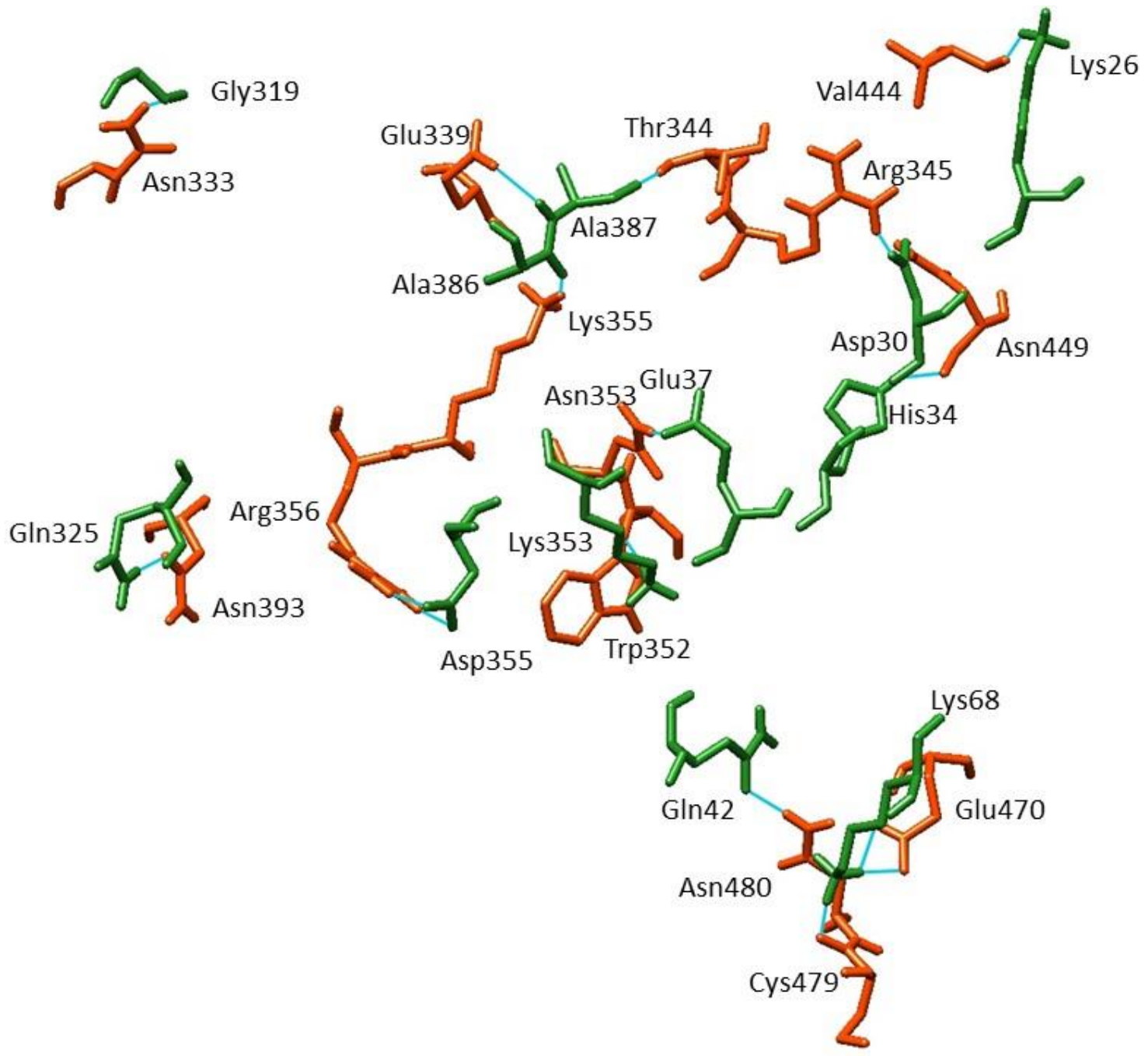

4(c)

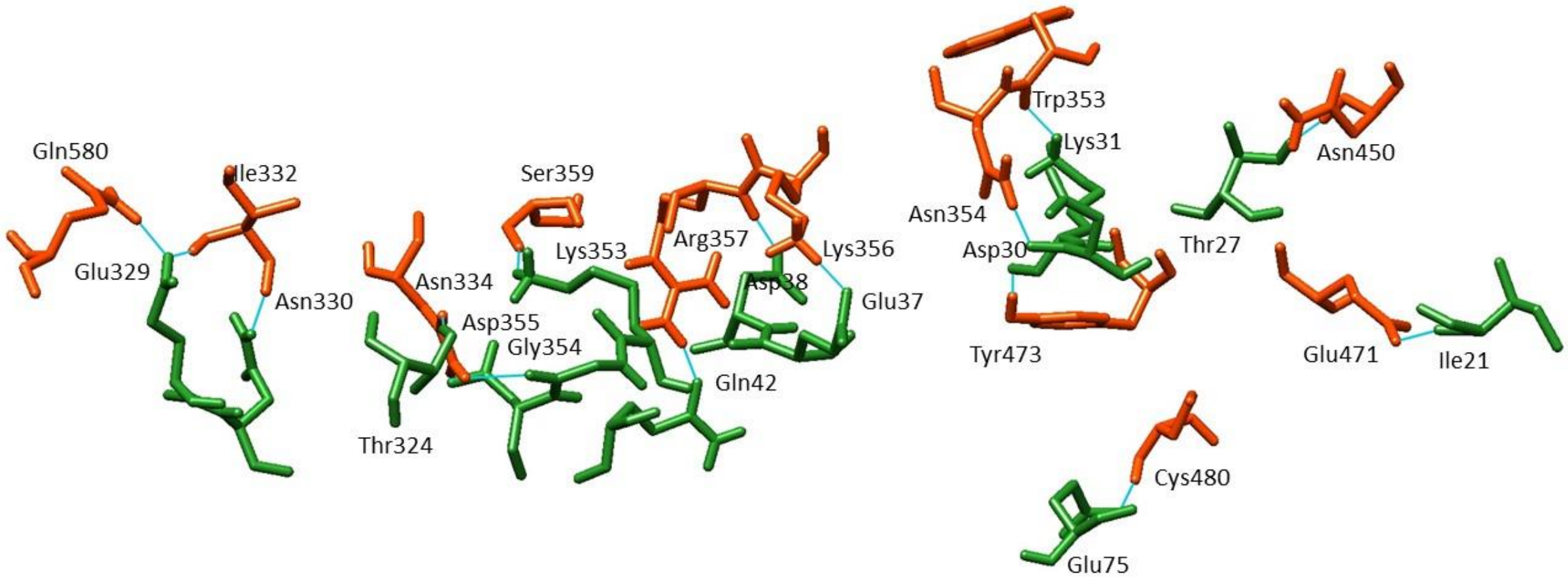

4(d)

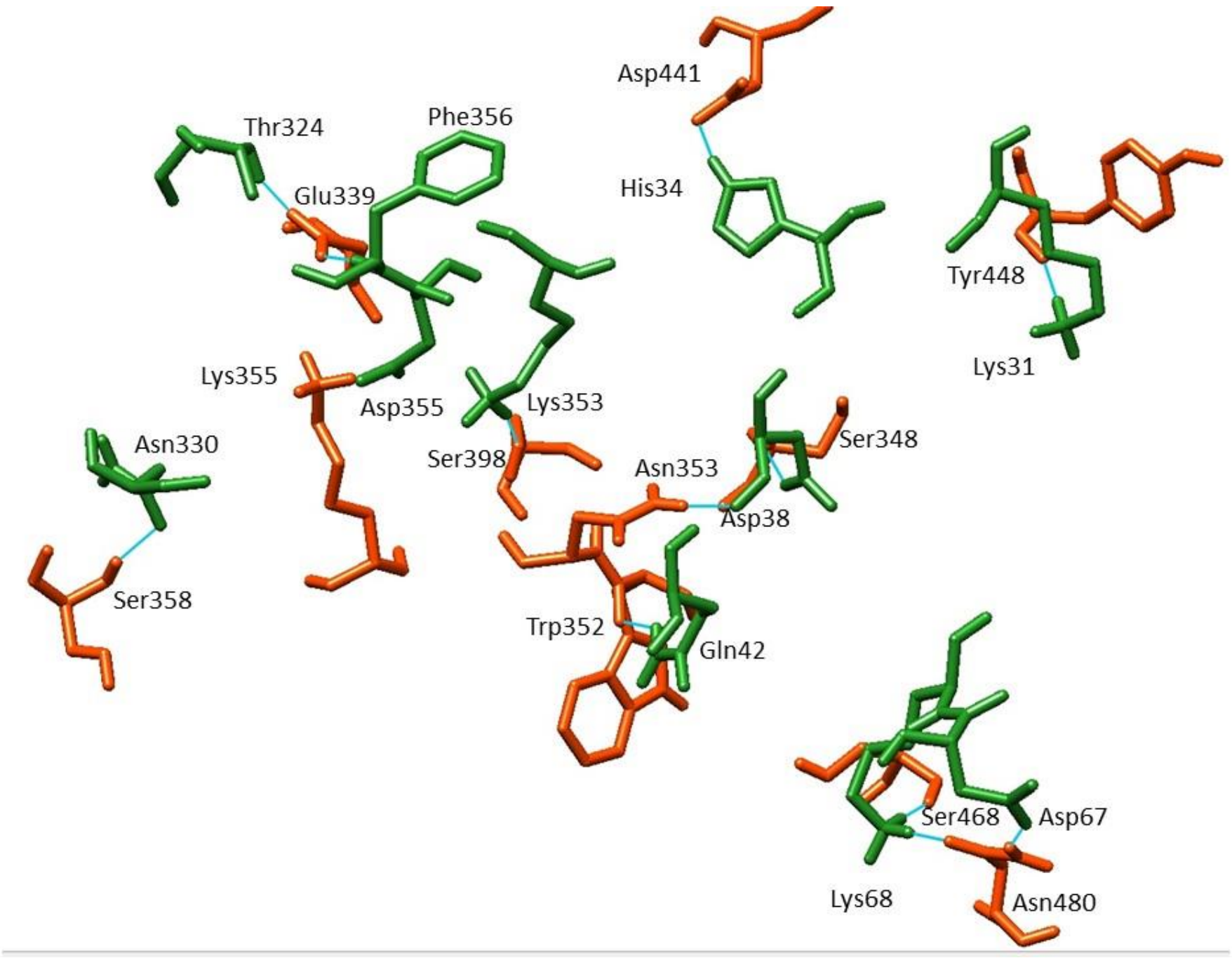

**Figure 4.** Protein-protein docking (a) SARS-CoV-2 Indian sequence 1 with complete ACE2, (b) SARS-CoV-2 Indian sequence 2 with complete ACE2, (c) Indian sequence 1 with soluble ACE2 and (d) Indian sequence 2 with soluble ACE2. The orange colour represents the Indian SARS-CoV-2 protein and dark green colour represents ACE2 protein.

5

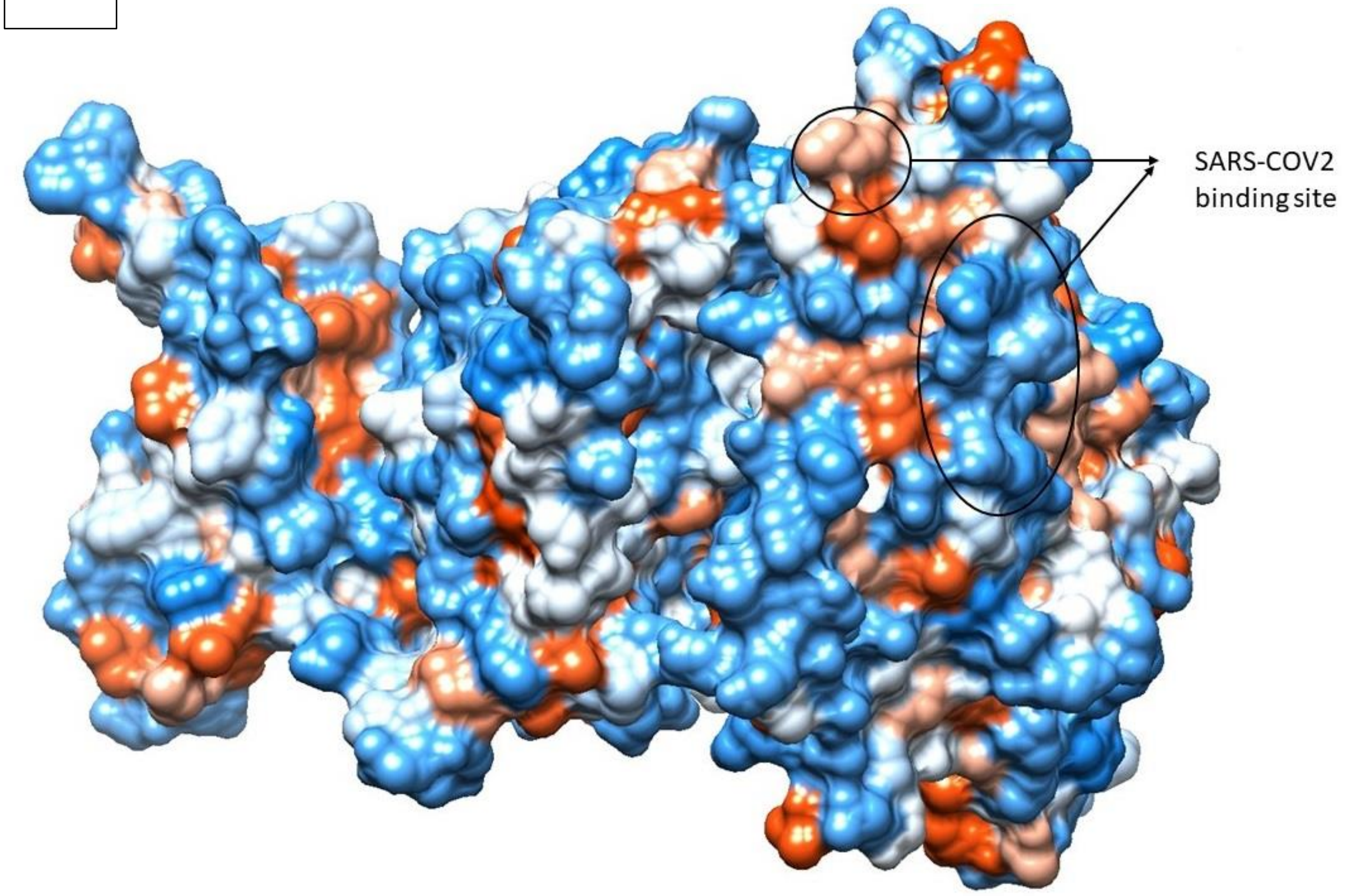

**Figure 5.** Surface view of ACE2 protein depicting the protein-protein binding site. The blue colour depicts the hydrophilic region and the red depicts the hydrophobic region.

6(a)

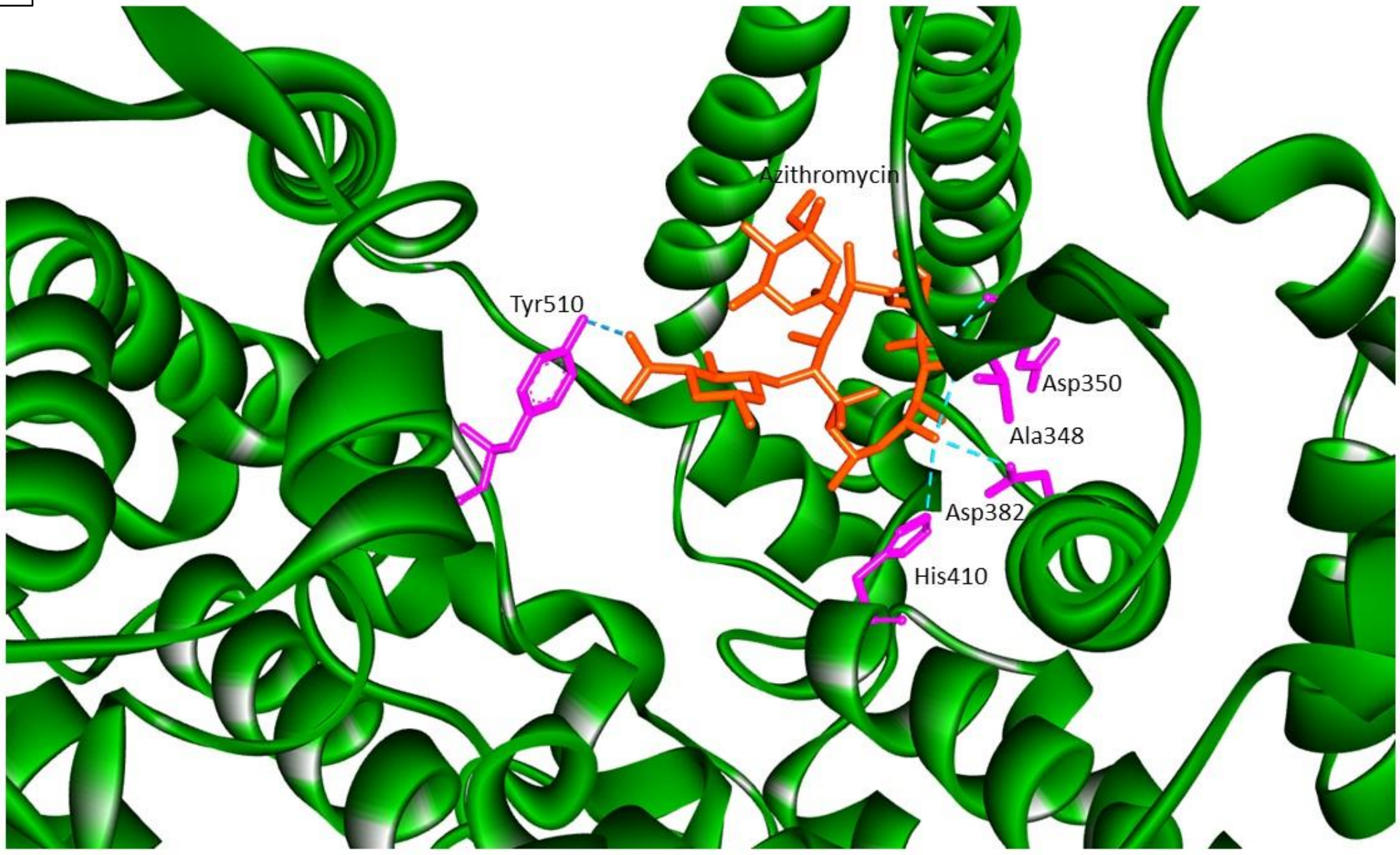

6(b)

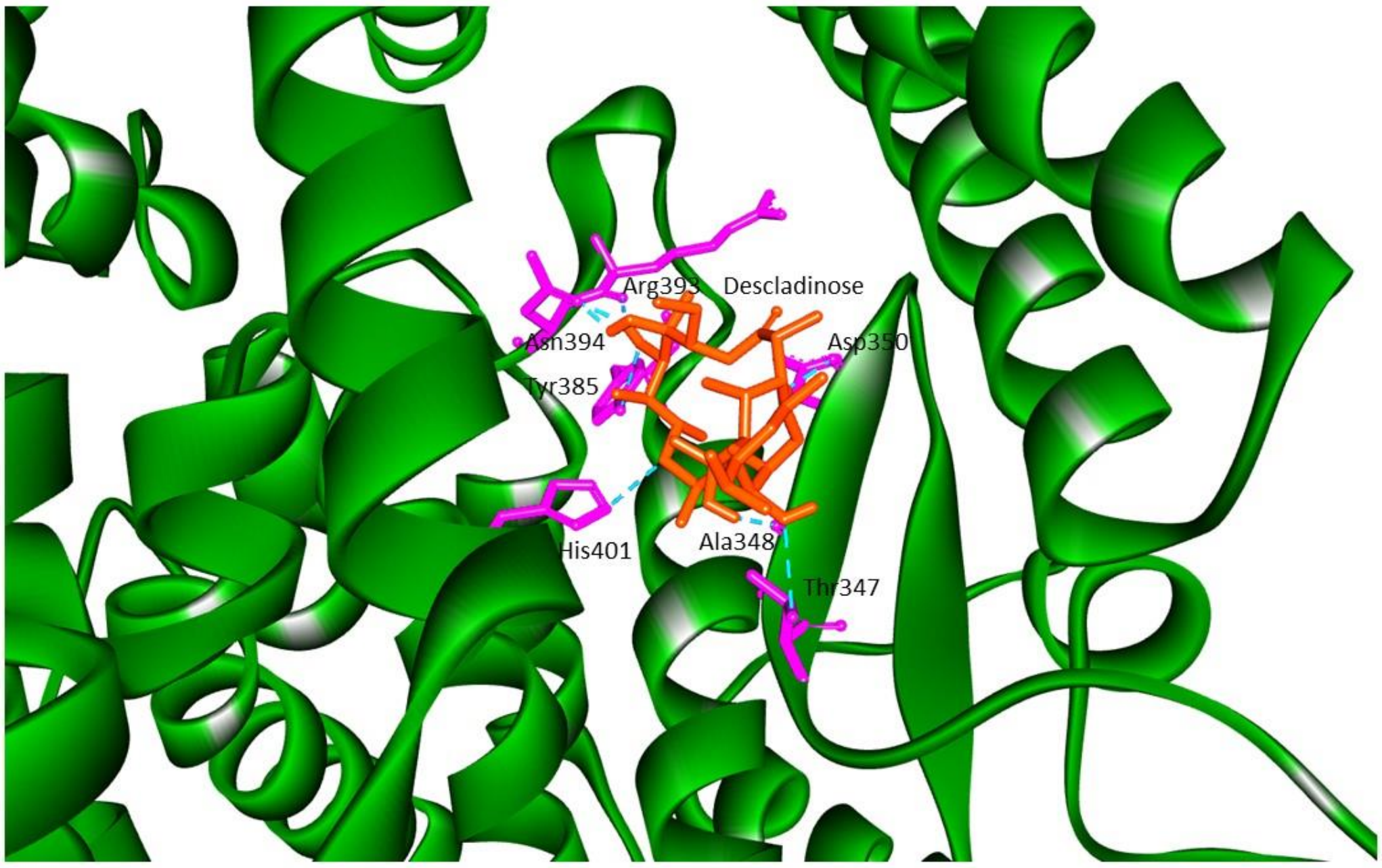

6(c)

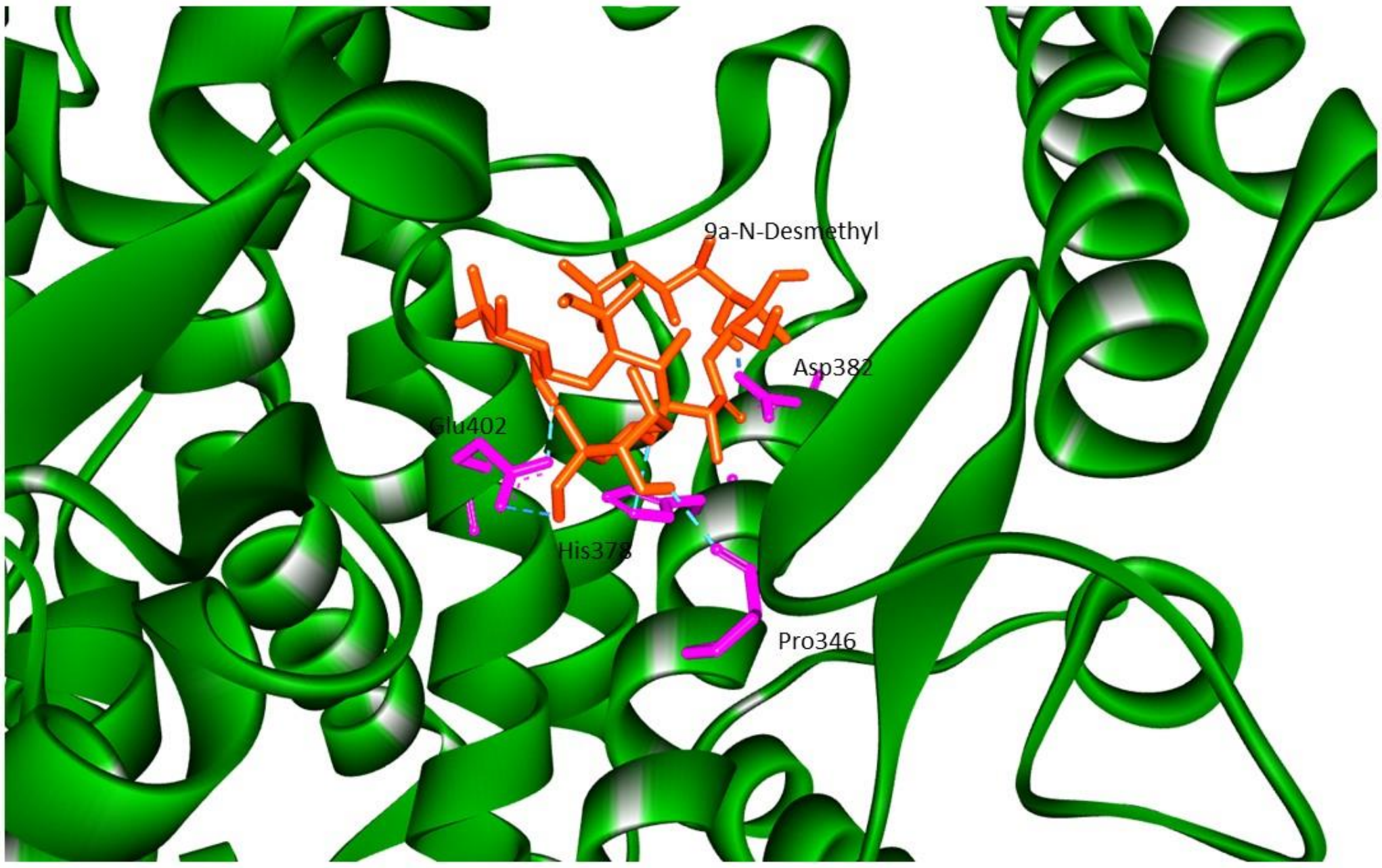
9a-N-Desmethyl
Asp382
Glu402
His378
Pro346

6(d)

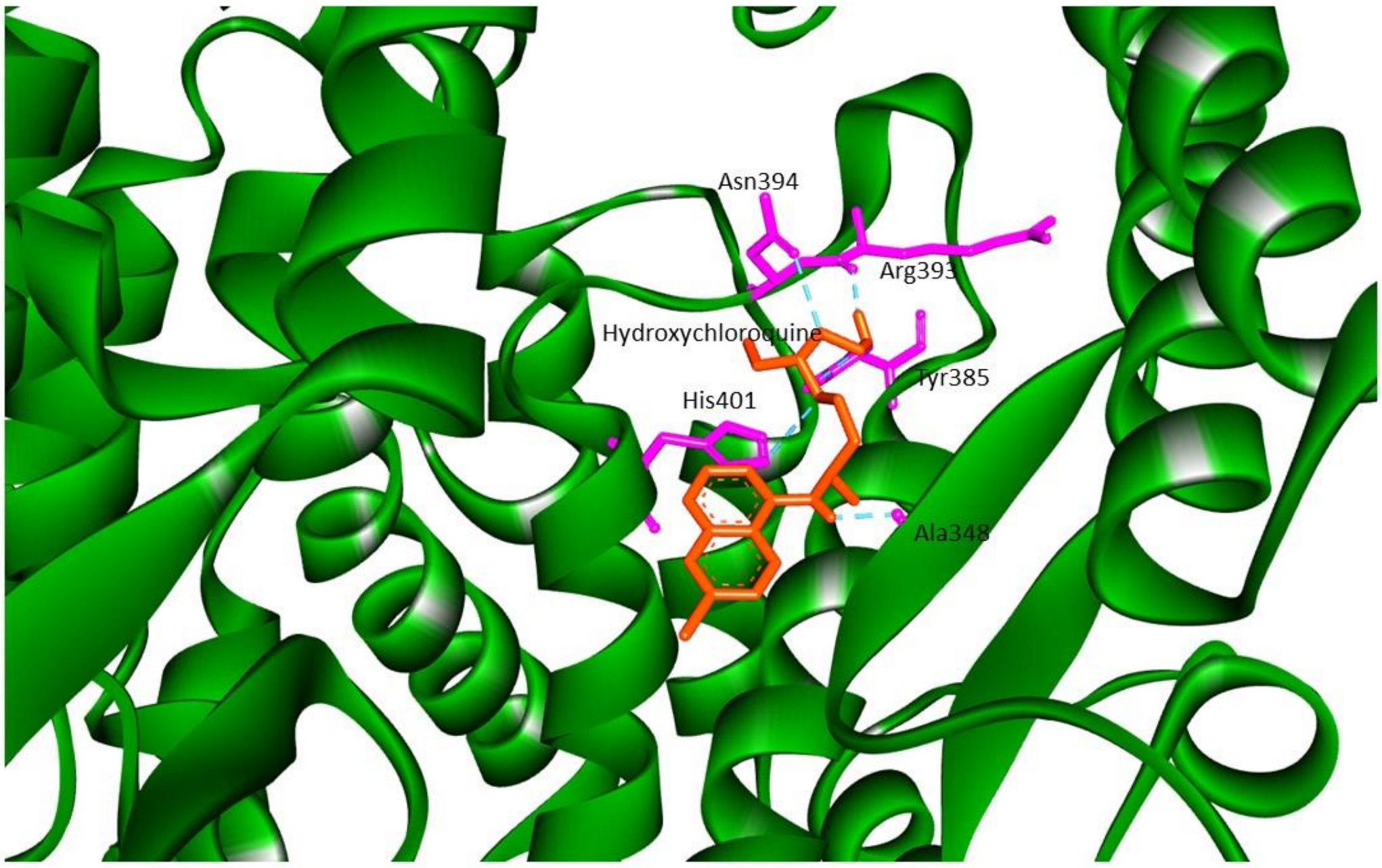

6(e)

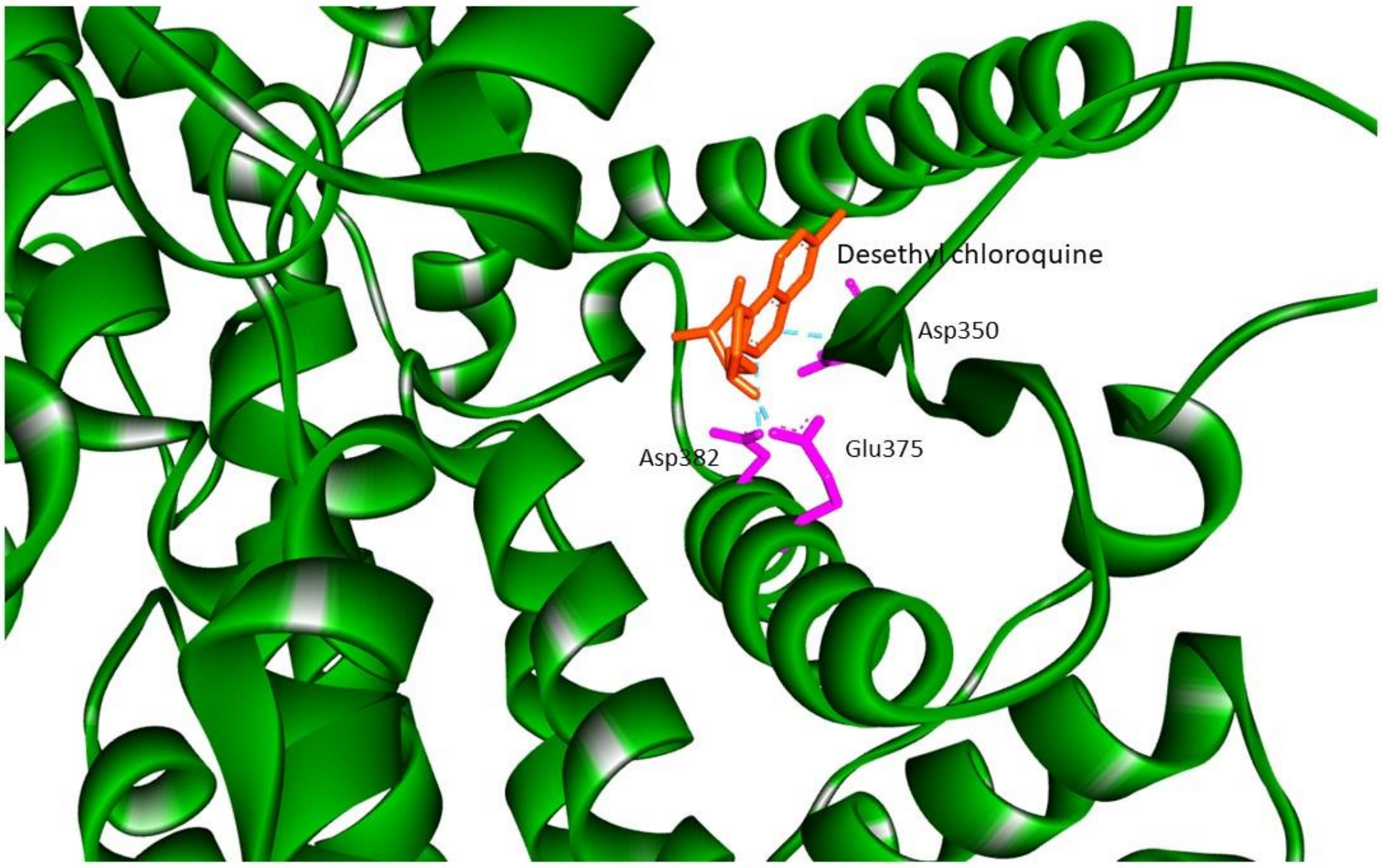
Desethyl chloroquine
Asp350
Asp382
Glu375

6(f)

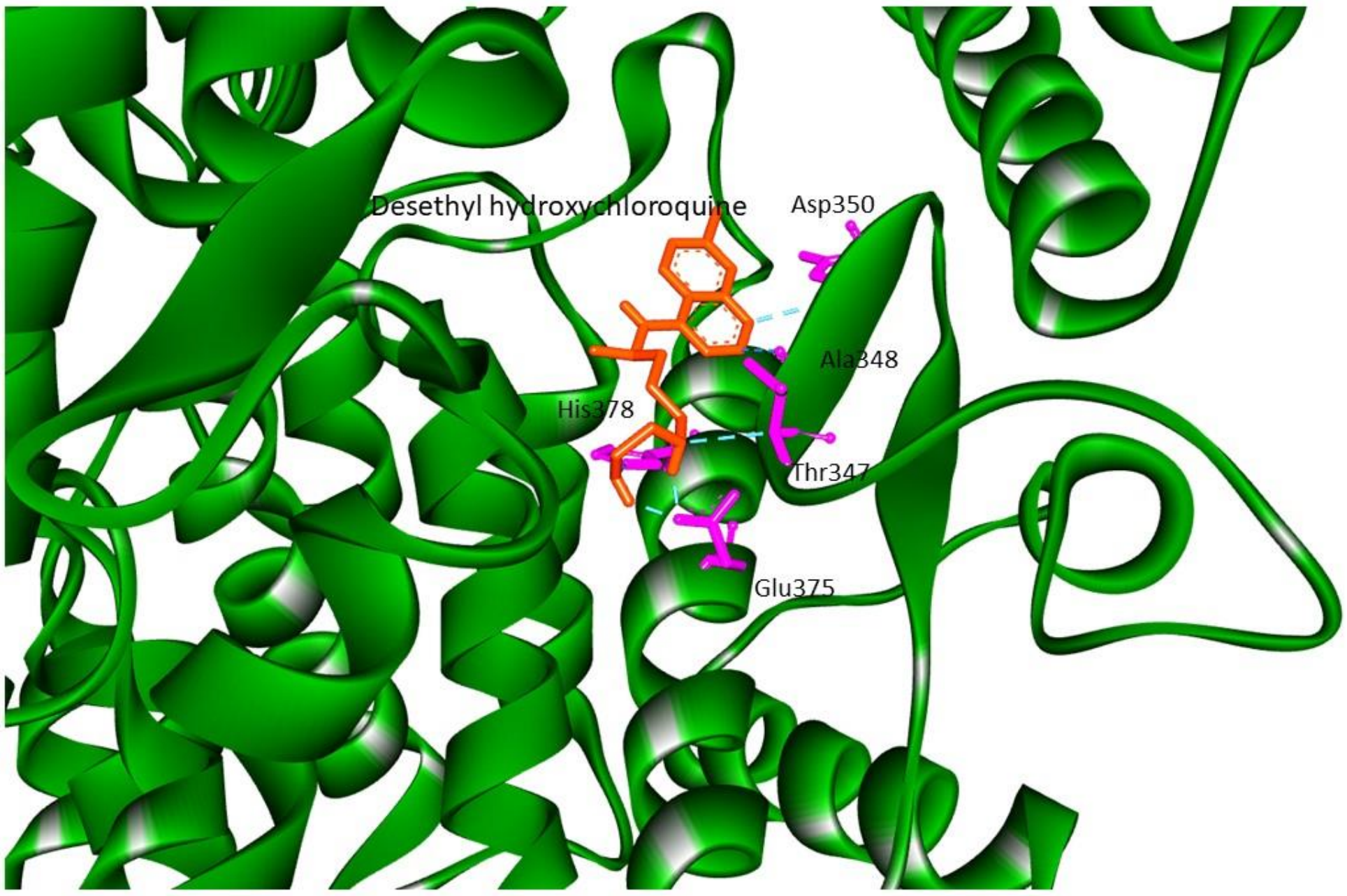

6(g)

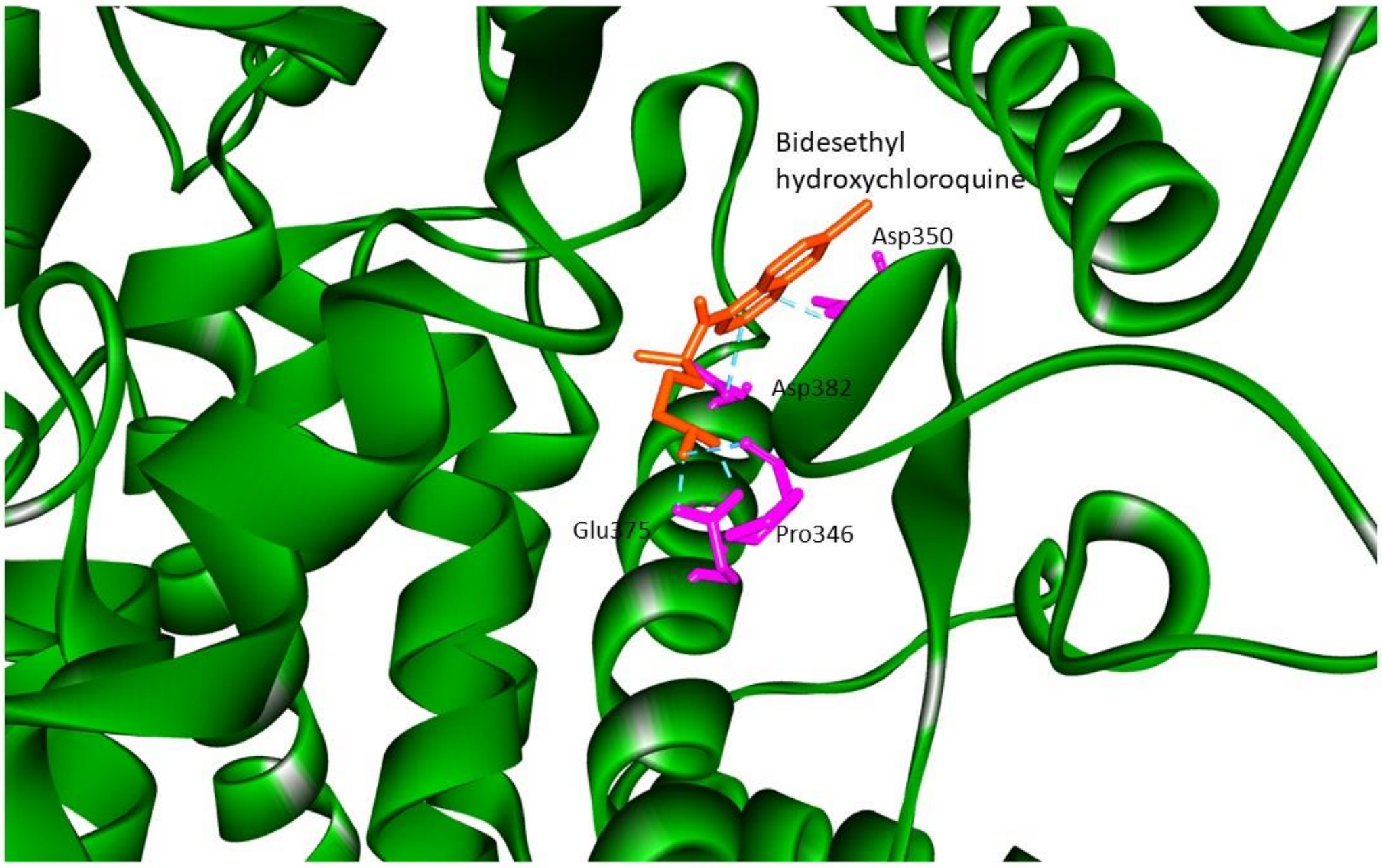

**Figure 6.** Protein-ligand docking of soluble ACE2 with (a) Azithromycin, (b) Descladinose-azithromycin, (c) 9a-N-Desmethyl-azithromycin, (d) Hydroxychloroquine, (e) Desethyl chloroquine, (f) Desethyl hydroxychloroquine, (g) Bisdesethyl hydroxychloroquine. The soluble ACE2 protein is shown in ribbon format in green colour, sidechains are shown in magenta and the drug in orange colour.

**Table 1. Protein-protein docking using HADDOCK online software wherein binding affinity was calculated using PRODIGY server.**

| SARS-CoV S1 -ACE2 interactions | HADDOCK values | | PRODIGY values kcal/mol |
|---|---|---|---|
| | HADDOCK score | Buried surface Area (Å²) | |
| **SARS-CoV with ACE2** | -165.1 +/-12.2 | 2896.2 +/- 450.1 | -10.3 |
| **INS1 (MT050493) with ACE2** | -178.6 +/-7.7 | 3706.2 +/- 130.8 | -18.1 |
| **INS2 (MT012098.1) with ACE2** | -112.3 +/-17.9 | 3475.5 +/- 361.5 | -14.7 |
| **INS1(MT050493) with sACE2** | -168.2 +/-8.7 | 3756.5 +/- 96.1 | -18.0 |
| **INS2(MT012098.1) with sACE2** | -121.1 +/-6.6 | 3160.1 +/-178.2 | -16.9 |

**Table 2. Interaction between INS 1 and 2 with ACE2 and sACE2.**

| Docking | Indian SARS-COV-2 S1 Sequences | ACE2 | Number of salt bridges |
|---|---|---|---|
| **INS1 with ACE2** | Ile332, Asn334, Asn354, Lys356, Arg357, Ser399, Cys480, Ala352, Trp353, Asp398, Thr470 | Glu329, Asp355, Phe356, Asp30, Asn33, Glu37, Asp38, Glu75, Lys31, His34, Tyr83, Asn330 | **3**<br>Lys356: Glu37<br>Arg357: Asp38<br>Asp398: His34 |
| **INS2 with ACE2** | Asn333, Thr344, Arg345, Asn353, Lys355, Arg356, Asn449, Asn480, Val444, Glu470, Cys479, Asn393, Trp352, Glu339 | Gly319, Ala387, Asp30, Glu37, Ala386, Asp355, Gln42, Lys26, His34, Lys68, Gln325, Lys353 | **3**<br>Arg345: Asp30<br>Arg356: Asp355<br>Glu470: Lys68 |
| **INS1 with sACE2** | Ile332, Asn334, Asn354, Lys356, Arg357, Tyr473, Cys480, Gln580, Glu471, Asn450, Trp353, Ser359 | Glu329, Gly354, Asp355, Asp30, Glu37, Asp38, Gln42, Lys31, Glu75, Ile21, Thr27, Thr324, Asn330, Lys353 | **2**<br>Lys356: Glu37<br>Arg357: Asp38 |
| **INS2 with sACE2** | Ser348, Asn353, Lys355, Ser358, Asn480, Tyr448, Asp441, Trp352, Ser468, Glu339, Ser398 | Asp38, Asp355, Asn330, Asp67, Lys31, His34, Gln42, Lys68, Thr324, Lys353, Phe356 | **2**<br>Lys355: Asp355<br>Asp441: His34 |

**Table 3. Protein-ligand docking for sACE2 interaction with Azithromycin, Hydroxychloroquine and their respective metabolites.**

| Chemical Compounds | Binding affinity (kcal/mol) | Inhibitory constant (µM) |
|---|---|---|
| **Azithromycin** | -7.7 | 2.0 |
| Descladinose | -8.03 | 1.3 |
| 9a-N-Desmethyl | -7.43 | 3.55 |
| **Hydroxychloroquine** | -5.75 | 60.78 |
| Desethyl chloroquine (DCQ) | -6.97 | 7.81 |
| Desethyl hydroxychloroquine | -6.71 | 12.13 |
| Bidesethyl hydroxychloroquine(BDCQ) | -7.62 | 2.62 |

To determine if the interaction was specific to the drugs, Pencillin structure was downloaded from PDB (id: PDB IFXV) and docked with sACE2. We observed a binding energy of -4.36 kcal/mol and an inhibitory constant value of 642.05 µM.

**Supplementary Figure 1.** Multiple sequence alignment of SARS-CoV-2 spike glycoprotein (S1) of SARS-CoV-NP_828851.1 (SARS-CoV-2003), Indian sequence 2 (MT012098.1), Indian sequence 1 (MT050493) and SARS-CoV2-WUHAN-QHD43416.

```
SARS-COV-NP_828851.1          MFIFLLFLTLTSGSDLDRCTTFDDVQAPNYTQHTSSMRGVYYPDEIFRSDTLYLTQDLFL      60
SARS-COV2-INS2                MFVFLVLLPLVSSQCVNLTT--RTQLPPAY--TNSFTRGVYYPDKVFRSSVLHSTQDLFL      56
SARS-COV2-INS1                MFVFLVLLPLVSSQCVNLTT--RTQLPPAY--TNSFTRGVYYPDKVFRSSVLHSTQDLFL      56
SARS-COV2-WUHAN-QHD43416      MFVFLVLLPLVSSQCVNLTT--RTQLPPAY--TNSFTRGVYYPDKVFRSSVLHSTQDLFL      56
                              **:**::* *.*.. ::  *       * *   .*  *******::***..*: ******

SARS-COV-NP_828851.1          PFYSNVTGFHTIN-------HTFGNPVIPFKDGIYFAATEKSNVVRGWVFGSTMNNKSQS      113
SARS-COV2-INS2                PFFSNVTWFHAIHVSGTNGTKRFDNPVLPFNDGVYFASTEKSNIIRGWIFGTTLDSKTQS      116
SARS-COV2-INS1                PFFSNVTWFHAIHVSGTNGTKRFDNPVLPFNDGVYFASTEKSNIIRGWIFGTTLDSKTQS      116
SARS-COV2-WUHAN-QHD43416      PFFSNVTWFHAIHVSGTNGTKRFDNPVLPFNDGVYFASTEKSNIIRGWIFGTTLDSKTQS      116
                              **:**** **:*:       : *.***:**:**:***:*****::***:**:*::.*:**

SARS-COV-NP_828851.1          VIIINNSTNVVIRACNFELCDNPFFAVSKPMGT----QTHTMIFDNAFNCTFEYISDAFS      169
SARS-COV2-INS2                LLIVNNATNVVIKVCEFQFCNDPFLGVY-HKNNKSWMESEFRVYSSANNCTFEYVSQPFL      175
SARS-COV2-INS1                LLIVNNATNVVIKVCEFQFCNDPFLGVYYHKNNKSWMESEFRVYSSANNCTFEYVSQPFL      176
SARS-COV2-WUHAN-QHD43416      LLIVNNATNVVIKVCEFQFCNDPFLGVYYHKNNKSWMESEFRVYSSANNCTFEYVSQPFL      176
                              ::*:**:*****:.*:*::*::**:.*    ..    ::.  ::..* ******:*: *

SARS-COV-NP_828851.1          LDVSEKSGNFKHLREFVFKNKDGFLYVYKGYQPIDVVRDLPSGFNTLKPIFKLPLGINIT      229
SARS-COV2-INS2                MDLEGKQGNFKNLREFVFKNIDGYFKIYSKHTPINLVRDLPQGFSALEPLVDLPIGINIT      235
SARS-COV2-INS1                MDLEGKQGNFKNLREFVFKNIDGYFKIYSKHTPINLVRDLPQGFSALEPLVDLPIGINIT      236
SARS-COV2-WUHAN-QHD43416      MDLEGKQGNFKNLREFVFKNIDGYFKIYSKHTPINLVRDLPQGFSALEPLVDLPIGINIT      236
                              :*:. *.****:******** **:: :*. : **::*****.**.:*:*:..**:*****

SARS-COV-NP_828851.1          NFRAILTAFS------PAQDIWGTSAAAYFVGYLKPTTFMLKYDENGTITDAVDCSQNPL      283
SARS-COV2-INS2                RFQTLLALHRSYLTPGDSSSGWTAGAAAYYVGYLQPRTFLLKYNENGTITDAVDCALDPL      295
SARS-COV2-INS1                RFQTLLALHRSYLTPGDSSSGWTAGAAAYYVGYLQPRTFLLKYNENGTITDAVDCALDPL      296
SARS-COV2-WUHAN-QHD43416      RFQTLLALHRSYLTPGDSSSGWTAGAAAYYVGYLQPRTFLLKYNENGTITDAVDCALDPL      296
                              .*:::*: .        :.. * :.****:****:* **:***:***********: :**

SARS-COV-NP_828851.1          AELKCSVKSFEIDKGIYQTSNFRVVPSGDVVRFPNITNLCPFGEVFNATKFPSVYAWERK      343
SARS-COV2-INS2                SETKCTLKSFTVEKGIYQTSNFRVQPTESIVRFPNITNLCPFGEVFNATRFASVYAWNRK      355
SARS-COV2-INS1                SETKCTLKSFTVEKGIYQTSNFRVQPTESIVRFPNITNLCPFGEVFNATRFASVYAWNRK      356
SARS-COV2-WUHAN-QHD43416      SETKCTLKSFTVEKGIYQTSNFRVQPTESIVRFPNITNLCPFGEVFNATRFASVYAWNRK      356
                              :* **::*** ::*********** *: .:*******************:* *****:**

SARS-COV-NP_828851.1          KISNCVADYSVLYNSTFFSTFKCYGVSATKLNDLCFSNVYADSFVVKGDDVRQIAPGQTG      403
SARS-COV2-INS2                RISNCVADYSVLYNSASFSTFKCYGVSPTKLNDLCFTNVYADSFVIRGDEVIQIAPGQTG      415
SARS-COV2-INS1                RISNCVADYSVLYNSASFSTFKCYGVSPTKLNDLCFTNVYADSFVIRGDEVRQIAPGQTG      416
SARS-COV2-WUHAN-QHD43416      RISNCVADYSVLYNSASFSTFKCYGVSPTKLNDLCFTNVYADSFVIRGDEVRQIAPGQTG      416
                              :**************: ********** ********:********::**:* ********
```

```
SARS-COV-NP_828851.1          VIADYNYKLPDDFMGCVLAWNTRNIDATSTGNYNYKYRYLRHGKLRPFERDISNVPFSPD    463
SARS-COV2-INS2                KIADYNYKLPDDFTGCVIAWNSNNLDSKVGGNYNYLYRLFRKSNLKPFERDISTEIYQAG    475
SARS-COV2-INS1                KIADYNYKLPDDFTGCVIAWNSNNLDSKVGGNYNYLYRLFRKSNLKPFERDISTEIYQAG    476
SARS-COV2-WUHAN-QHD43416      KIADYNYKLPDDFTGCVIAWNSNNLDSKVGGNYNYLYRLFRKSNLKPFERDISTEIYQAG    476
                               ************ ***:***:.*:*:.  ***** ** :*:.:*:*******.  :. .

SARS-COV-NP_828851.1          GKPCTP-PALNCYWPLNDYGFYTTTGIGYQPYRVVVLSFELLNAPATVCGPKLSTDLIKN    522
SARS-COV2-INS2                STPCNGVEGFNCYFPLQSYGFQPTNGVGYQPYRVVVLSFELLHAPATVCGPKKSTNLVKN    535
SARS-COV2-INS1                STPCNGVEGFNCYFPLQSYGFQPTNGVGYQPYRVVVLSFELLHAPATVCGPKKSTNLVKN    536
SARS-COV2-WUHAN-QHD43416      STPCNGVEGFNCYFPLQSYGFQPTNGVGYQPYRVVVLSFELLHAPATVCGPKKSTNLVKN    536
                              ..**.   .:***:**:.***  *.*:***************:********* **:*:**

SARS-COV-NP_828851.1          QCVNFNFNGLTGTGVLTPSSKRFQPFQQFGRDVSDFTDSVRDPKTSEILDISPCAFGGVS    582
SARS-COV2-INS2                KCVNFNFNGLTGTGVLTESNKKFLPFQQFGRDIADTTDAVRDPQTLEILDITPCSFGGVS    595
SARS-COV2-INS1                KCVNFNFNGLTGTGVLTESNKKFLPFQQFGRDIADTTDAVRDPQTLEILDITPCSFGGVS    596
SARS-COV2-WUHAN-QHD43416      KCVNFNFNGLTGTGVLTESNKKFLPFQQFGRDIADTTDAVRDPQTLEILDITPCSFGGVS    596
                              :**************** *.*:* ********::* **:****:* *****:**:*****

SARS-COV-NP_828851.1          VITPGTNASSEVAVLYQDVNCTDVSTAIHADQLTPAWRIYSTGNNVFQTQAGCLIGAEHV    642
SARS-COV2-INS2                VITPGTNTSNQVAVLYQDVNCTEVPVAIHADQLTPTWRVYSTGSNVFQTRAGCLIGAEHV    655
SARS-COV2-INS1                VITPGTNTSNQVAVLYQDVNCTEVPVAIHADQLTPTWRVYSTGSNVFQTRAGCLIGAEHV    656
SARS-COV2-WUHAN-QHD43416      VITPGTNTSNQVAVLYQDVNCTEVPVAIHADQLTPTWRVYSTGSNVFQTRAGCLIGAEHV    656
                              *******:*.:***********:* .*********:**:****.*****:**********

SARS-COV-NP_828851.1          DTSYECDIPIGAGICASYHTVSL----LRSTSQKSIVAYTMSLGADSSIAYSNNTIAIPT    698
SARS-COV2-INS2                NNSYECDIPIGAGICASYQTQTNSPRRARSVASQSIIAYTMSLGAENSVAYSNNSIAIPT    715
SARS-COV2-INS1                NNSYECDIPIGAGICASYQTQTNSPRRARSVASQSIIAYTMSLGAENSVAYSNNSIAIPT    716
SARS-COV2-WUHAN-QHD43416      NNSYECDIPIGAGICASYQTQTNSPRRARSVASQSIIAYTMSLGAENSVAYSNNSIAIPT    716
                              :.****************:* :      **.:.:**:********:.*:*****:*****

SARS-COV-NP_828851.1          NFSISITTEVMPVSMAKTSVDCNMYICGDSTECANLLLQYGSFCTQLNRALSGIAAEQDR    758
SARS-COV2-INS2                NFTISVTTEILPVSMTKTSVDCTMYICGDSTECSNLLLQYGSFCTQLNRALTGIAVEQDK    775
SARS-COV2-INS1                NFTISVTTEILPVSMTKTSVDCTMYICGDSTECSNLLLQYGSFCTQLNRALTGIAVEQDK    776
SARS-COV2-WUHAN-QHD43416      NFTISVTTEILPVSMTKTSVDCTMYICGDSTECSNLLLQYGSFCTQLNRALTGIAVEQDK    776
                              **:**:***::****:******.**********:*****************:***.***:

SARS-COV-NP_828851.1          NTREVFAQVKQMYKTPTLKYFGGFNFSQILPDPLKPTKRSFIEDLLFNKVTLADAGFMKQ    818
SARS-COV2-INS2                NTQEVFAQVKQIYKTPPIKDFGGFNFSQILPDPSKPSKRSFIEDLLFNKVTLADAGFIKQ    835
SARS-COV2-INS1                NTQEVFAQVKQIYKTPPIKDFGGFNFSQILPDPSKPSKRSFIEDLLFNKVTLADAGFIKQ    836
SARS-COV2-WUHAN-QHD43416      NTQEVFAQVKQIYKTPPIKDFGGFNFSQILPDPSKPSKRSFIEDLLFNKVTLADAGFIKQ    836
                              **:********:**** :* ************* **:********************:**

SARS-COV-NP_828851.1          YGECLGDINARDLICAQKFNGLTVLPPLLTDDMIAAYTAALVSGTATAGWTFGAGAALQI    878
SARS-COV2-INS2                YGDCLGDIAARDLICAQKFNGLTVLPPLLTDEMIAQYTSALLAGTITSGWTFGAGAALQI    895
SARS-COV2-INS1                YGDCLGDIAARDLICAQKFNGLTVLPPLLTDEMIAQYTSALLAGTITSGWTFGAGAALQI    896
SARS-COV2-WUHAN-QHD43416      YGDCLGDIAARDLICAQKFNGLTVLPPLLTDEMIAQYTSALLAGTITSGWTFGAGAALQI    896
                              **:***** **********************:*** **:**::** *:************
```

```
SARS-COV-NP_828851.1          YGECLGDINARDLICAQKFNGLTVLPPLTDDMIAAYTAALVSGTATAGWTFGAGAALQI    878
SARS-COV2-INS2                YGDCLGDIAARDLICAQKFNGLTVLPPLTDEMIAQYTSALLAGTITSGWTFGAGAALQI    895
SARS-COV2-INS1                YGDCLGDIAARDLICAQKFNGLTVLPPLTDEMIAQYTSALLAGTITSGWTFGAGAALQI    896
SARS-COV2-WUHAN-QHD43416      YGDCLGDIAARDLICAQKFNGLTVLPPLTDEMIAQYTSALLAGTITSGWTFGAGAALQI    896
                              **:***** *********************:*** **:**::** *:************

SARS-COV-NP_828851.1          PFAMQMAYRFNGIGVTQNVLYENQKQIANQFNKAISQIQESLTTTSTALGKLQDVVNQNA   938
SARS-COV2-INS2                PFAMQMAYRFNGIGVTQNVLYENQKLIANQFNSAIGKIQDSLSSTASALGKLQDVVNQNA   955
SARS-COV2-INS1                PFAMQMAYRFNGIGVTQNVLYENQKLIANQFNSVIGKIQDSLSSTASALGKLQDVVNQNA   956
SARS-COV2-WUHAN-QHD43416      PFAMQMAYRFNGIGVTQNVLYENQKLIANQFNSAIGKIQDSLSSTASALGKLQDVVNQNA   956
                              ************************* ******..*.:**:**::*::*************

SARS-COV-NP_828851.1          QALNTLVKQLSSNFGAISSVLNDILSRLDKVEAEVQIDRLITGRLQSLQTYVTQQLIRAA   998
SARS-COV2-INS2                QALNTLVKQLSSNFGAISSVLNDILSRLDKVEAEVQIDRLITGRLQSLQTYVTQQLIRAA   1015
SARS-COV2-INS1                QALNTLVKQLSSNFGAISSVLNDILSRLDKVEAEVQIDRLITGRLQSLQTYVTQQLIRAA   1016
SARS-COV2-WUHAN-QHD43416      QALNTLVKQLSSNFGAISSVLNDILSRLDKVEAEVQIDRLITGRLQSLQTYVTQQLIRAA   1016
                              ************************************************************

SARS-COV-NP_828851.1          EIRASANLAATKMSECVLGQSKRVDFCGKGYHLMSFPQAAPHGVVFLHVTYVPSQERNFT   1058
SARS-COV2-INS2                EIRASANLAATKMSECVLGQSKRVDFCGKGYHLMSFPQSAPHGVVFLHVTYVPAQEKNFT   1075
SARS-COV2-INS1                EIRASANLAATKMSECVLGQSKRVDFCGKGYHLMSFPQSAPHGVVFLHVTYVPAQEKNFT   1076
SARS-COV2-WUHAN-QHD43416      EIRASANLAATKMSECVLGQSKRVDFCGKGYHLMSFPQSAPHGVVFLHVTYVPAQEKNFT   1076
                              **************************************:**************:**:***

SARS-COV-NP_828851.1          TAPAICHEGKAYFPREGVFVFNGTSWFITQRNFFSPQIITTDNTFVSGNCDVVIGIINNT   1118
SARS-COV2-INS2                TAPAICHDGKAHFPREGVFVSNGTHWFVTQRNFYEPQIITTDNTFVSGNCDVVIGIVNNT   1135
SARS-COV2-INS1                TAPAICHDGKAHFPREGVFVSNGTHWFVTQRNFYEPQIITTDNTFVSGNCDVVIGIVNNT   1136
SARS-COV2-WUHAN-QHD43416      TAPAICHDGKAHFPREGVFVSNGTHWFVTQRNFYEPQIITTDNTFVSGNCDVVIGIVNNT   1136
                              *******:***:******** *** **:*****:.*********************:***

SARS-COV-NP_828851.1          VYDPLQPELDSFKEELDKYFKNHTSPDVDLGDISGINASVVNIQKEIDRLNEVAKNLNES   1178
SARS-COV2-INS2                VYDPLQPELDSFKEELDKYFKNHTSPDVDLGDISGINASVVNIQKEIDRLNEVAKNLNES   1195
SARS-COV2-INS1                VYDPLQPELDSFKEELDKYFKNHTSPDVDLGDISGINASVVNIQKEIDRLNEVAKNLNES   1196
SARS-COV2-WUHAN-QHD43416      VYDPLQPELDSFKEELDKYFKNHTSPDVDLGDISGINASVVNIQKEIDRLNEVAKNLNES   1196
                              ************************************************************

SARS-COV-NP_828851.1          LIDLQELGKYEQYIKWPWYVWLGFIAGLIAIVMVTILLCCMTSCCSCLKGACSCGSCCKF   1238
SARS-COV2-INS2                LIDLQELGKYEQYIKWPWYIWLGFIAGLIAIVMVTIMLCCMTSCCSCLKGCCSCGSCCKF   1255
SARS-COV2-INS1                LIDLQELGKYEQYIKWPWYIWLGFIAGLIAIVMVTIMLCCMTSCCSCLKGCCSCGSCCKF   1256
SARS-COV2-WUHAN-QHD43416      LIDLQELGKYEQYIKWPWYIWLGFIAGLIAIVMVTIMLCCMTSCCSCLKGCCSCGSCCKF   1256
                              *******************:****************:*************.*********

SARS-COV-NP_828851.1          DEDDSEPVLKGVKLHYT 1255
SARS-COV2-INS2                DEDDSEPVLKGVKLHYT 1272
SARS-COV2-INS1                DEDDSEPVLKGVKLHYT 1273
SARS-COV2-WUHAN-QHD43416      DEDDSEPVLKGVKLHYT 1273
                              *****************
```